\documentclass[a4paper,11pt]{article}

\usepackage{jheppub}
\usepackage{amsmath,amssymb,amsthm,mathtools,bm,booktabs,tabularx}
\usepackage{tikz}
\usetikzlibrary{decorations.pathmorphing}

\usepackage{float}
\usepackage{hyperref}
\hypersetup{%
    colorlinks,%
    breaklinks,%
    citecolor=red,urlcolor=brown,linkcolor=blue,%
    hypertexnames=false,%
    pdftitle={The Kerr Metric in Canonical Harmonic Coordinates to All Orders},%
    pdfauthor={A. Sadu, P. Vanhove}%
}

\newcommand{\GN}{G_{\!N}}
\newcommand{\gth}{\mathfrak{g}} 
\newcommand{\hth}{\mathfrak{h}}
\newcommand{\bx}{\mathbf{x}}
\newcommand{\bk}{\mathbf{k}} 
\newcommand{\ba}{\mathbf{a}}
\newcommand{\bS}{\mathbf{S}}
\newcommand{\bn}{\mathbf{n}}
\newcommand{\zh}{\hat{\mathbf z}}
\newcommand{\ephi}{\hat{\mathbf e}_\varphi}
\newcommand{\evarpi}{\hat{\mathbf e}_\varpi}
\newcommand{\Lin}{\mathrm{lin}}
\newcommand{\cE}{\mathcal{E}}
\newcommand{\cD}{\mathcal{D}}

\newtheorem{lemma}{Lemma}
\newtheorem{theorem}{Theorem}

\theoremstyle{remark}

\tikzset{grav/.style={decorate,decoration={snake,amplitude=1.1pt,segment length=4.5pt}},
         aux/.style={dashed}}
\newcommand{\vertexJJ}[3]{%
\begin{tikzpicture}[baseline=-2pt,scale=0.85]
\draw[#1] (0,0)--(0,0.75); \draw[#2] (0,0)--(-0.55,-0.6); \draw[#3] (0,0)--(0.55,-0.6);
\end{tikzpicture}}
\newcommand{\lineG}{\tikz[baseline=-2.5pt]\draw[grav](0,0)--(0.9,0);}
\newcommand{\lineY}{\tikz[baseline=-2.5pt]\draw[aux](0,0)--(0.9,0);}
\newcommand{\vertexS}[3]{%
\begin{tikzpicture}[baseline=-2.5pt,scale=0.8]
\draw[#1] (-1.0,0)--(0,0); \draw[#2] (0,0)--(0.8,0.42); \draw[#3] (0,0)--(0.8,-0.42);
\end{tikzpicture}}

\title{The Kerr Metric in Canonical Harmonic Coordinates to All Orders}

\author[a]{Amina Sadu}
\author[b,c]{and Pierre Vanhove}
\affiliation[a]{Universit\'e Paris-Saclay, F-91405 Orsay, France}
\affiliation[b]{Institut de Physique Th\'eorique, Universit\'e Paris-Saclay, CEA, CNRS,\\ F-91191 Gif-sur-Yvette Cedex, France}
\affiliation[c]{Laboratoire d’Annecy-le-Vieux de Physique Th\'eorique (LAPTh), CNRS and Universit\'e Savoie Mont-Blanc, Annecy, 74940, France}
\preprint{LAPTH-061/26}

\abstract{ We give in closed form the transformation from Boyer--Lindquist
    coordinates to the canonical harmonic coordinates of the Kerr metric, to all
    orders in Newton's constant and in the spin. This is the canonical harmonic
    chart selected by the MPM prescription and by the harmonic-gauge Einstein
    recursion sourced by the Kerr multipoles, with the residual homogeneous
    freedom fixed by the boundary conditions. The chart is the real,
    time-reversal-covariant combination of the two radial harmonic functions
    underlying the Cook--Scheel coordinates, which differ from it by a definite
    residual harmonic transformation. The proof rests on a classification of the
    residual harmonic freedom. Rotational covariance, parity, time reversal and
    analyticity in the spin allow a decaying homogeneous solution beyond the
    linear field only in the spatial components at third post-Minkowskian order,
    with angular degree $\ell=p+2$ at spin order $a^p$. This sector is a pure
    gauge mode, equivalent to an MPM gauge moment. 
    The exact chart selects the representative in which this third-PM
    homogeneous mode vanishes; imposing the same condition on the recursion
    removes the only residual ambiguity and fixes its solution uniquely.
    As a consequence, the canonical MPM iteration needs no finite-part
    regularisation and no harmonicity correction at any order. We give the exact
    gothic metric and the post-Minkowskian coefficients in closed form through
    fifth order, where an arctangent function first appears at fourth order. We
    find agreement with recent canonical MPM and amplitude-based results at
    fourth order, and we verify the construction in exact arithmetic through
    fifth order and twelfth order in the spin, with code made publicly
    available. }

\keywords{Black Holes, Classical Theories of Gravity, Scattering Amplitudes}

\begin{document}
\maketitle
\flushbottom

\section{Introduction}\label{sec:intro}

Harmonic coordinates are the working coordinates of post-Newtonian and
post-Minkowskian gravity. In the multipolar post-Minkowskian (MPM)
formalism~\cite{Blanchet:1985sp,Blanchet:2013haa} the exterior field of an
isolated source is built in \emph{canonical} harmonic coordinates, specified by
the source multipole moments, with all gauge moments set to zero, and by a
definite inverse-Laplacian prescription. Amplitude-based reconstructions of
classical metrics~\cite{Duff:1973zz,Bjerrum-Bohr:2002ks,Neill:2013wsa,Mougiakakos:2020laz,Jakobsen:2020ksu,Damgaard:2024fqj,Mougiakakos:2024nku}
also produce the metric in harmonic coordinates, because the de~Donder
condition on the gothic inverse metric $\gth^{\mu\nu}=\sqrt{-g}\,g^{\mu\nu}$,
$\partial_\mu\gth^{\mu\nu}=0$, is the exact harmonic-coordinate condition. For
the Schwarzschild metric the harmonic chart is elementary, and the amplitude
recursion was solved to all orders in Newton's constant using the cubic
formulation of Einstein gravity~\cite{Mougiakakos:2024nku,Cheung:2017kzx}; a
related perturbiner construction uses Landau--Lifshitz variables and an
auxiliary inverse metric~\cite{Damgaard:2024fqj}.

For the Kerr metric the situation is less clear. Harmonicity does not fix the
coordinates: different harmonic charts are related by residual harmonic
transformations, and several harmonic forms of the Kerr metric are
known~\cite{Ruiz:1986kerr,Cook:1997qc,Liu:1998kerr,Lin:2014laa,Zang:2020kerr}.
The most widely used, the Cook--Scheel chart~\cite{Cook:1997qc} on which the
closed-form metric of~\cite{Lin:2014laa} is based, is not the one produced by
perturbation theory. In the perturbiner approach
of~\cite{Damgaard:2026star,Damgaard:2026kerr} it is reached only by adding a
gauge term at order $\GN^2$, and the representative obtained without that term
has been evaluated through fourth post-Minkowskian (PM) order, where its
coefficients are no longer rational. The canonical MPM metric of Kerr has
likewise been computed through fourth PM order, to all orders in the
spin~\cite{Xing:2026rrd}. There, the absence of finite parts in the inverse
Laplacian and the vanishing of the harmonicity correction were verified
through fifth PM order and assumed beyond, and the transformation from
Boyer--Lindquist to canonical harmonic coordinates was obtained only through
third order. A closed form valid to all orders, and an explanation of why
the perturbative representative takes the form it does, have so far been
missing.

In this paper we give that closed form. We show that the harmonic-gauge
Einstein recursion sourced by the Kerr multipoles, solved without adding
decaying homogeneous solutions, and the canonical MPM iteration both produce
the Kerr metric in one and the same harmonic chart, to all orders in $\GN$ and
in the spin, and we give that chart explicitly. The proof rests on a
classification of the homogeneous solutions that the symmetries of a
stationary spinning source allow, which reduces the residual freedom to a
single gauge sector at third PM order.

\medskip
Our main results are:

\begin{enumerate}
\item \textit{Canonical harmonic coordinates of Kerr in closed form.} With
$R(r):=r-\GN M$, the chart
\begin{equation}\label{eq:mainchart}
T=t,\qquad
x+iy=\operatorname{Re}\!\left[(R(r)+ia)e^{-i\Theta(r)}\right]\sin\theta\,e^{i\phi},
\qquad z=R(r)\cos\theta,
\end{equation}
where
\begin{equation}\label{eq:Theta}
\Theta(r)=\int_r^\infty\frac{a\,du}{u^2-2\GN Mu+a^2}
=\begin{cases}
\displaystyle \frac{a}{2q}\log\frac{R(r)+q}{R(r)-q},
& |a|<\GN M,\quad q=\sqrt{\GN^2M^2-a^2},\\[6pt]
\displaystyle \frac{a}{R(r)},& |a|=\GN M,\\[6pt]
\displaystyle \frac{a}{\chi}\arctan\frac{\chi}{R(r)},
& |a|>\GN M,\quad \chi=\sqrt{a^2-\GN^2M^2}.
\end{cases}
\end{equation}
is the chart selected by the recursion, and it coincides to all PM orders with
the canonical harmonic chart of the MPM formalism. Eq.~\eqref{eq:mainchart} is
therefore the transformation from Boyer--Lindquist to canonical harmonic
coordinates. For a subextremal black hole the logarithmic branch applies in the
exterior region $R>q$, where the integral defines it unambiguously, and it
covers the extremal limit; the $\arctan$ branch is the superextremal analytic
continuation. The chart is the real, time-reversal-covariant combination of the
two complex-conjugate radial harmonic functions underlying the Cook--Scheel
chart and its time reverse, and the Cook--Scheel chart differs from it by a
definite residual harmonic transformation, given explicitly in terms of the
Boyer--Lindquist radius in sections~\ref{sec:threecharts} and~\ref{sec:canonical}.

\item \textit{Classification of the residual harmonic freedom.} Rotational
covariance, parity, time reversal and analyticity in the spin imply that,
beyond the linear field, a decaying homogeneous solution can occur only in the
spatial components at order $\GN^3$, with angular degree $\ell=p+2$ at spin
order $a^p$ (Theorem~\ref{thm:homog}). This sector is spanned by a pure gauge
mode, which corresponds to an MPM gauge moment. The chart~\eqref{eq:mainchart}
has no component in it, and imposing the same condition on the recursion fixes
the solution uniquely to all orders (Theorems~\ref{thm:unique}
and~\ref{thm:chart}). The absence of logarithms follows order by order, since
the exact chart supplies a solution of the same Poisson equations whose
expansion contains none. If time reversal is not imposed, additional
homogeneous terms are allowed at order $\GN^2$; the Cook--Scheel chart is an
example. The Schwarzschild problem has a direct antecedent: there a
one-parameter residual gauge freedom also first appears at third PM
order~\cite{Mougiakakos:2020laz}. The classification extends that observation
to a rotating source and determines the allowed spin-dependent angular
structures.

\item \textit{Consequences for the canonical MPM metric.} The identification
(Theorem~\ref{thm:canonical}) establishes to all orders the properties of the
canonical Kerr metric that were checked or assumed in~\cite{Xing:2026rrd}:
the iteration meets no resonance, so no finite part is needed; the
harmonicity correction vanishes at every order. It also gives in closed form
the transformation from Boyer--Lindquist to canonical coordinates, and
explicitly the relation between canonical and Cook--Scheel coordinates.

\item \textit{Explicit metric and checks.} We give the exact gothic metric in
this chart and the PM coefficients in closed form through fifth order. They are
polynomials of degree at most two in a single transcendental function, which
contains $\arctan(a/\varrho)$, where $\varrho$ is the flat oblate spheroidal
radius defined in \eqref{eq:oblate}, enters from fourth order. They agree
with the canonical MPM metric of~\cite{Xing:2026rrd} and with the
gauge-term-free result of~\cite{Damgaard:2026kerr} at fourth order. We verify
the construction in exact arithmetic through $\GN^5$ and spin order $a^{12}$,
with an independent Landau--Lifshitz computation and an iteration from the
linear field alone through $\GN^8$. The code is publicly
available~\cite{Sadu:2026code}.
\end{enumerate}

\paragraph{Hypotheses.} The all-order statement is a uniqueness result with
explicit hypotheses. The exact Kerr geometry supplies a candidate harmonic chart,
and the stationary harmonic-gauge Einstein equations supply the perturbative problem.
That problem is specified by the Kerr multipoles, which fix the linear field, by
asymptotic flatness, by the symmetry conditions (i)--(iii) of
section~\ref{sec:structure}, and by the boundary prescription in the exceptional
third-order sector. We prove that the two constructions agree order by order.
The source enters only through the linear field, obtained from the spinning
three-point amplitude, whose spin dependence
exponentiates~\cite{Arkani-Hamed:2017jhn,Guevara:2018wpp,Chung:2018kqs,Arkani-Hamed:2019ymq}
and reproduces the linearised Kerr field~\cite{Vines:2017hyw,Chung:2019yfs},
with multipoles $\mathcal M_\ell+i\mathcal S_\ell=M(ia)^\ell$~\cite{Geroch:1970cd,Hansen:1974zz}.
The assumptions behind this point-source prescription are discussed in
section~\ref{sec:nonlinwl}.

\paragraph{Related work.} The recursive de~Donder-gauge exterior metrics
of~\cite{Damgaard:2026star} for general stationary mass and current multipoles
are explicit through second PM order at arbitrary multipole order and recover
Kerr for a special choice of multipoles. They clarify the distinction between
specifying exterior multipole data and deriving those data from a source action,
but they do not give an all-order Kerr metric or select a chart.
In~\cite{BjerrumBohr:2026polyakov} the Kerr multipole source and its
one-graviton coupling are derived from a Polyakov-type hypersurface action,
and~\cite{BjerrumBohr:2026resum} studies the resummation of the spin of the
heavy body in aligned-spin scattering at third PM order. These works concern
the source and the observables, whereas the present paper concerns the
stationary spacetime and its coordinates.

\medskip
The paper is organised as follows. Section~\ref{sec:cubic} sets up the
harmonic-gauge field equations and their recursion, using the cubic formulation
as a computational representation. Section~\ref{sec:source} constructs the Kerr
source from the spinning-particle amplitude and the worldline action. Section~\ref{sec:structure}
classifies the homogeneous solutions and proves the uniqueness result.
Section~\ref{sec:chart} constructs the exact harmonic chart and identifies it
with the chart selected by the recursion, with the Cook--Scheel chart up to a
residual transformation, and with the canonical MPM chart.
Section~\ref{sec:PM} gives the closed-form PM expansion through fifth order,
and section~\ref{sec:checks} collects the checks. We conclude in
section~\ref{sec:discussion}. Appendix~\ref{app:LL} gives the Landau--Lifshitz
form of the relaxed equations. Appendix~\ref{app:tangherlini} discusses the
harmonic radius of the Schwarzschild--Tangherlini metric in $D$ dimensions,
where the third-order resonance appears as a pole in $D-4$ and the limit
$D\to4$ gives the standard harmonic chart. Appendix~\ref{app:coefficients}
lists the lowest-order coefficients of the double expansion in $\GN$ and in
the spin. Appendix~\ref{app:cubic} collects the rewriting of the action in the
cubic formulation, its Feynman rules and the derivation of the field equations.

\section{Einstein equations, source and perturbative recursion}\label{sec:cubic}

The metric of a massive body is the classical one-point function of the
gravitational field in the presence of the body. It can be obtained either from
tree diagrams with one external graviton or, equivalently, by solving the
classical Einstein equations with the body as a
source~\cite{Duff:1973zz,Mougiakakos:2020laz,Mougiakakos:2024nku,Damgaard:2024fqj,Damgaard:2026kerr}.
Our classification and all-order identification are statements about these
harmonic-gauge Einstein equations and their boundary data, and they do not
depend on a particular action formulation: the structural results of
sections~\ref{sec:structure} and~\ref{sec:chart} use only the harmonic-gauge
field equations, the symmetries of the source and the inverse Laplacian of
section~\ref{sec:propagator}. For practical calculations we use the cubic
formulation of gravity~\cite{Cheung:2017kzx,Mougiakakos:2024nku}, because it
gives the most economical recursion, with a non-propagating auxiliary field and
cubic interactions only, and because it makes its interpretation in terms of
tree amplitudes manifest. The Landau--Lifshitz relaxed equations, given in
appendix~\ref{app:LL}, provide an independent formulation and check. In this
section we give only what the recursion requires. The rewriting of the action,
its Feynman rules and the derivation of the field equations are collected in
appendix~\ref{app:cubic}.

\subsection{Field equations in cubic form}

The cubic formulation of~\cite{Cheung:2017kzx} uses the gothic inverse metric
\begin{equation}\label{eq:gothic}
    \gth^{\mu\nu}:=\sqrt{-g}\,g^{\mu\nu}\,,\qquad g:=\det(g_{\mu\nu})\,,
\end{equation}
and a symmetric auxiliary field $A^\lambda{}_{\mu\nu}=A^\lambda{}_{\nu\mu}$.
With the de~Donder gauge-fixing term of~\cite{Mougiakakos:2024nku} the
Einstein-Hilbert Lagrangian reads
\begin{equation}\label{eq:action}
16\pi\GN\,\mathcal L_{\rm EH}=-Q(A,A;\gth)+A^\lambda{}_{\mu\nu}\,\partial_\lambda\gth^{\mu\nu}-\frac12\,\eta_{\rho\sigma}\,\partial_\mu\gth^{\mu\rho}\,\partial_\nu\gth^{\nu\sigma}\,,
\end{equation}
where we introduced the bilinear form
\begin{equation}\label{eq:Qdef}
Q(A,B;\gth):=\Big(A^\lambda{}_{\mu\rho}B^\mu{}_{\lambda\sigma}-\frac{1}{D-1}\,A_\rho B_\sigma\Big)\gth^{\rho\sigma}\,,\qquad A_\sigma:= A^\lambda{}_{\lambda\sigma}\,,
\end{equation}
which is symmetric in $A$ and $B$. The Lagrangian is cubic in $(A,\gth)$, and
since $\partial^\lambda\eta_{\lambda\mu}=0$ the gauge-fixing term is exactly
quadratic in $\hth^{\mu\nu}=\eta^{\mu\nu}-\gth^{\mu\nu}$. The field equation
of the auxiliary field $A$ is algebraic,
\begin{equation}
    {\delta \mathcal L_{\rm EH}\over \delta A^\lambda{}_{\mu\nu}}
    =-\frac{1}{16\pi \GN}\,\left(\cE(A;\gth)_\lambda{}^{\mu\nu}-\partial_\lambda\gth^{\mu\nu}\right)
    =0\,,
\end{equation}
where we have set
\begin{equation}\label{eq:Aeq}
\cE(A;\gth)_\lambda{}^{\mu\nu}:=  A^\mu{}_{\lambda\sigma}\gth^{\nu\sigma}
+A^\nu{}_{\lambda\sigma}\gth^{\mu\sigma}-\frac{1}{D-1}\big(\delta^\mu_\lambda A_\sigma\gth^{\nu\sigma}
+\delta^\nu_\lambda A_\sigma\gth^{\mu\sigma}\big)\,,
\end{equation}
by the identity~\eqref{eq:QE} of appendix~\ref{app:cubic}. The field equation
of $\gth^{\mu\nu}$ is, outside the source,
\begin{equation}\label{eq:geq}
{\delta \mathcal L_{\rm EH}\over \delta \gth^{\mu\nu}}= \partial_\lambda A^\lambda{}_{\mu\nu}+(AA)_{\mu\nu}-\eta_{\sigma(\mu}\partial_{\nu)}\partial_\rho\gth^{\rho\sigma}=0\,,
\end{equation}
where we have used the notation
\begin{equation}
     (AA)_{\mu\nu}:= A^\rho{}_{\sigma\mu}A^\sigma{}_{\rho\nu}
     -\frac{1}{D-1}\,A_\mu A_\nu=\frac{\partial Q(A,A;\gth)}{\partial\gth^{\mu\nu}}\,.
\end{equation}
On shell, the solution of~\eqref{eq:Aeq} is expressed through the Christoffel
symbols of $g_{\mu\nu}=|g|^{-1/2}\gth_{\mu\nu}$,
\begin{equation}\label{eq:Aonshell}
A^\lambda{}_{\mu\nu}=-\Gamma^\lambda_{\mu\nu}
+\frac12\left(\delta^\lambda_\mu\,\partial_\nu+\delta^\lambda_\nu\,\partial_\mu\right)\log\sqrt{-g}\,,
\end{equation}
which follows from writing the first-order (Palatini) form of the
Einstein--Hilbert action in terms of the shifted connection
$-\big(\Gamma^\lambda_{\mu\nu}-\delta^\lambda_{(\mu}\Gamma^\rho_{\nu)\rho}\big)$~\cite{Cheung:2017kzx}.
Since $\gth^{\mu\nu}$ is a tensor density, the de~Donder condition
$\partial_\mu\gth^{\mu\nu}=0$ is the exact harmonic coordinate condition $\Box_g
x^\nu=0$, and the perturbative metric can be compared with exact metrics in
harmonic coordinates~\cite{Fromholz:2013hka,Mougiakakos:2024nku}.

\subsection{Variables, propagators and source}\label{sec:Seff}

The constant linear map $\cE_0:=\cE(\,\cdot\,;\eta)$ is invertible for $D>2$.
Lowering all indices with $\eta_{\mu\nu}$, its inverse is
\begin{equation}\label{eq:E0inv}
\big(\cE_0^{-1}X\big)_{\lambda\mu\nu}=\frac12\big(X_{\mu\lambda\nu}+X_{\nu\lambda\mu}-X_{\lambda\mu\nu}\big)+\frac{1}{2(D-2)}\,\eta_{\mu\nu}\,X_{\lambda\rho}{}^{\rho}\,,
\end{equation}
the Christoffel-symbol combination with a trace correction. The solution of the
linearised equation~\eqref{eq:Aeq}, $\cE_0(A)_{\lambda}{}^{\mu\nu}=-\partial_\lambda\hth^{\mu\nu}$, is therefore
\begin{equation}\label{eq:Alin}
A_{\Lin}[\hth]^\lambda{}_{\mu\nu}=-\frac12\,\eta^{\lambda\sigma}\left(\partial_\mu\hth_{\nu\sigma}+\partial_\nu\hth_{\mu\sigma}-\partial_\sigma\hth_{\mu\nu}+\frac{1}{D-2}\,\eta_{\mu\nu}\partial_\sigma\hth\right).
\end{equation}
Following~\cite[eq.~(6)]{Mougiakakos:2024nku}, we shift the auxiliary field by
this linearised on-shell value,
\begin{equation}\label{eq:Ydef}
A^\lambda{}_{\mu\nu}=A_{\Lin}[\hth]^\lambda{}_{\mu\nu}+Y^\lambda{}_{\mu\nu}\,.
\end{equation}
In the variables $(\hth,Y)$ the quadratic part of the action is diagonal
(appendix~\ref{app:cubic}, eq.~\eqref{eq:Lbulk}). The graviton has the de~Donder
kinetic term, written with the trace reversal
\begin{equation}\label{eq:Pdef}
(PX)_{\mu\nu}:= X_{\mu\nu}-\frac{1}{D-2}\,\eta_{\mu\nu}\,\eta^{\rho\sigma}X_{\rho\sigma}\,,\qquad (P^{-1}X)_{\mu\nu}=X_{\mu\nu}-\frac12\,\eta_{\mu\nu}\,\eta^{\rho\sigma}X_{\rho\sigma}\,,
\end{equation}
whereas the quadratic term of $Y$ has no derivatives, so that $Y$ does not
propagate. The interactions consist of three cubic vertices, $\hth\hth\hth$,
$\hth\hth Y$ and $\hth YY$, with two, one and zero derivatives, and there are
no quartic or higher vertices (eq.~\eqref{eq:vertices} and
table~\ref{tab:rules}). For a source term $\int d^Dx\,\hth^{\mu\nu}j_{\mu\nu}$,
the static field is
\begin{equation}\label{eq:props}
\hth_{\mu\nu}(\bk)=\frac{32\pi\GN}{\bk^2}\,\big(P^{-1}j\big)_{\mu\nu}(\bk)\,.
\end{equation}
The numerator $P^{-1}$ of the graviton propagator is independent of $D$;
see~\cite{Mougiakakos:2024nku} for a discussion of the dimension dependence of
the propagator and of the physical observables. The body couples to the metric
only, not to the auxiliary field. To linear order in $\hth$ its coupling is the
source $j=\frac12PT$, where $T^{\mu\nu}$ is its stress tensor in flat space
(section~\ref{sec:source}, eq.~\eqref{eq:Swlexp}).

\paragraph{Power counting.} Each propagator carries a factor $\GN$, each vertex
a factor $1/\GN$ and each insertion of the source a factor $M$. A tree graph
with $n$ insertions of the source and one external graviton has $n-1$ cubic
vertices and $2n-1$ propagators, so it is of order $\GN^nM^n$. The order $\GN^n$
of the post-Minkowskian expansion is the sum of the trees with $n$ insertions of
the source.

\subsection{Field equations and the classical recursion}

In the variables $(\hth,Y)$ the field equations are
(appendix~\ref{app:cubic})
\begin{equation}\label{eq:Yeq}
\cE_0(Y)=\cE(A;\hth)\,,
\end{equation}
which is~\eqref{eq:Aeq} written in terms of $Y$, and
\begin{equation}\label{eq:heq}
\frac12\,\Box(P\hth)_{\mu\nu}+(AA)_{\mu\nu}+\partial_\lambda Y^\lambda{}_{\mu\nu}=-8\pi\GN\,(PT)_{\mu\nu}\,,
\end{equation}
which is the field equation~\eqref{eq:geq} of $\gth$ in the presence of the
source. Its linear part, $\Box\,\hth^{(1)\mu\nu}=-16\pi\GN\,T^{\mu\nu}$, gives
the first-order field of section~\ref{sec:source}.

We expand $\hth^{\mu\nu}=\sum_{n\ge1}\hth^{(n)\mu\nu}$,
$A^\lambda{}_{\mu\nu}=\sum_{n\ge1}A^{(n)\lambda}{}_{\mu\nu}$ and
$Y^\lambda{}_{\mu\nu}=\sum_{n\ge2}Y^{(n)\lambda}{}_{\mu\nu}$, where the
superscript $(n)$ is the order in $\GN$. The momentum-space currents of the
graviton and of the auxiliary field are
\begin{equation}
    \begin{aligned}
\hth^{(n)\mu\nu}(\bx)&=\int\!\frac{d^{D-1}k}{(2\pi)^{D-1}}\,e^{i\bk\cdot\bx}\,J^{(n)\mu\nu}(\bk)\,,\cr
Y^{(n)\lambda}{}_{\mu\nu}(\bx)&=\int\!\frac{d^{D-1}k}{(2\pi)^{D-1}}\,e^{i\bk\cdot\bx}\,\tilde Y^{(n)\lambda}{}_{\mu\nu}(\bk)\,.
    \end{aligned}
\end{equation}
In the notation of~\cite{Mougiakakos:2024nku}, where
$\gth^{\mu\nu}=\eta^{\mu\nu}-\sqrt{32\pi\GN}\,h^{\mu\nu}$, their $h$ is
$\hth/\sqrt{32\pi\GN}$ and their auxiliary field, after the shift of
\cite[eq.~(6)]{Mougiakakos:2024nku}, is $Y/\sqrt{32\pi\GN}$. Up to the overall normalization and coupling factors factored out in~\cite{Mougiakakos:2024nku}, their currents $J^{(n)}$ and $Y^{(n)}$ of~\cite[eqs.~(14), (15)]{Mougiakakos:2024nku} correspond to $J^{(n)}$ and $\widetilde{Y}^{(n)}$ here.

For a static source $\partial_0=0$, so that $\Box=\Delta$. In $D=4$ and
outside the source, eqs.~\eqref{eq:Yeq} and~\eqref{eq:heq} become recursion
relations, solved order by order in $\GN$. At order $n$, the auxiliary field
$Y^{(n)}=A^{(n)}-A_{\Lin}[\hth^{(n)}]$ is determined algebraically by the
lower orders,
\begin{equation}\label{eq:recY}
Y^{(n)}=\cE_0^{-1}\Big[\sum_{m=1}^{n-1}\cE\big(A^{(m)};\hth^{(n-m)}\big)\Big],
\end{equation}
and in particular $Y^{(1)}=0$. The field $\hth^{(n)}$ then solves the Poisson
equation
\begin{equation}\label{eq:recJ}
\frac12\,\Delta\Big(\hth^{(n)}_{\mu\nu}-\frac12\,\eta_{\mu\nu}\hth^{(n)}\Big)
=-\sum_{m=1}^{n-1}\Big(A^{(m)\rho}{}_{\sigma\mu}A^{(n-m)\sigma}{}_{\rho\nu}
-\frac13A^{(m)}_\mu A^{(n-m)}_\nu\Big)
-\partial_\lambda Y^{(n)\lambda}{}_{\mu\nu}\,,
\end{equation}
whose source involves $Y^{(n)}$ and the fields of order lower than $n$. It is solved with the inverse Laplacian~\eqref{eq:invlap} in section~\ref{sec:propagator}, followed by applying the inverse trace-reversal $P^{-1}$ and raising indices with $\eta^{\mu\nu}$ to obtain $\hth^{(n)\mu\nu}$.

\begin{figure}[t]
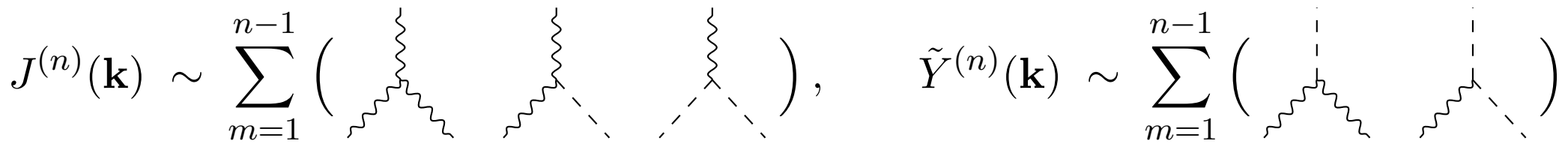

\centering
$\displaystyle
J^{(n)}(\bk)\;\sim\;\sum_{m=1}^{n-1}\Big(\;\vertexJJ{grav}{grav}{grav}\;\;\;\;\vertexJJ{grav}{grav}{aux}\;\;\;\;\vertexJJ{grav}{aux}{aux}\;\Big)\,,\qquad
\tilde Y^{(n)}(\bk)\;\sim\;\sum_{m=1}^{n-1}\Big(\;\vertexJJ{aux}{grav}{grav}\;\;\;\;\vertexJJ{aux}{grav}{aux}\;\Big)$
\caption{Structure of the recursion~\eqref{eq:recY}--\eqref{eq:recJ}: the
graviton current $J^{(n)}(\bk)$ (wavy) and the auxiliary current $\tilde Y^{(n)}(\bk)$ (dashed)
at order $\GN^n$ are built from the three cubic vertices~\eqref{eq:vertices}
acting on currents of orders $m$ and $n-m$, see~\cite{Mougiakakos:2024nku}. The
outgoing line is on top. The lowest-order current $J^{(1)}(\bk)$ is the one-graviton
emission by the source, and $\tilde Y^{(1)}(\bk)=0$.}
\label{fig:recursion}
\end{figure}

Each term has a diagrammatic meaning, shown in figure~\ref{fig:recursion}.
In~\eqref{eq:recY}, the terms with $A^{(m)}\to A_{\Lin}[\hth^{(m)}]$ come from
the $\hth\hth Y$ vertex and those with $A^{(m)}\to Y^{(m)}$ from the $\hth YY$
vertex, with the outgoing auxiliary line attached through the contact propagator
$\cE_0^{-1}$. In~\eqref{eq:recJ}, the products $A^{(m)}A^{(n-m)}$ are the three
vertices with the outgoing graviton attached to their undifferentiated leg. The
term $\partial_\lambda Y^{(n)\lambda}{}_{\mu\nu}$ collects the graphs in which
the outgoing graviton is attached instead to a differentiated leg
$A_{\Lin}[\hth]$ of the $\hth\hth\hth$ and $\hth\hth Y$ vertices and by the
symmetry of $\cE_0^{-1}$ they combine into the divergence of the auxiliary
current of the same order. Only the one-graviton emission $\hth^{(1)}$ refers to
the source. In the terminology of~\cite{Damgaard:2024fqj},
eqs.~\eqref{eq:recY}--\eqref{eq:recJ} are the perturbiner recursion of the
action~\eqref{eq:Lbulk}, with the currents $J^{(n)}$ and $\tilde Y^{(n)}$ as
perturbiner coefficients. The non-propagating field $Y$ plays a role similar to
that of the auxiliary inverse metric introduced
in~\cite{Damgaard:2024fqj,Damgaard:2026kerr} to avoid an infinite tower of
vertices so that here all interactions are cubic.

\subsection{The static propagator in position space}\label{sec:propagator}

For a static source the recursion is most conveniently solved in position space.
Every term generated at order $\GN^n$ is a sum of terms
$H_\ell(\bx)/|\bx|^{\mu}$, where $H_\ell(\bx)$ is a homogeneous harmonic polynomial
of degree $\ell$. The inverse Laplacian acts as
\begin{equation}\label{eq:invlap}
\Delta^{-1}\,\frac{H_\ell(\bx)}{|\bx|^{\mu}}=\frac{1}{(\mu-2)(\mu-2\ell-3)}\,\frac{H_\ell(\bx)}{|\bx|^{\mu-2}}\,,
\end{equation}
which is the position-space version of the Fourier integrals of the
recursion~\cite{Mougiakakos:2024nku} given by
\begin{equation}\label{eq:FT}
\int_{\mathbb R^3} d^3x\;e^{-i\bk\cdot\bx}\,\frac{H_\ell(\bx)}{|\bx|^{\mu}}=(-i)^\ell\,2^{3+\ell-\mu}\pi^{\frac32}\,\frac{\Gamma\big(\frac{3+2\ell-\mu}{2}\big)}{\Gamma\big(\frac{\mu}{2}\big)}\,\frac{H_\ell(\bk)}{|\bk|^{3+2\ell-\mu}}\,,
\end{equation}
the $\Gamma$-function factors are those of the expansion parameter $\rho(|\bk|,D,n)$
of~\cite{Mougiakakos:2024nku} evaluated at $D=4$. 

\medskip
We stress two features of~\eqref{eq:invlap} that will be of importance later.

\begin{itemize}
\item \emph{Homogeneous solutions.} The harmonic functions
$H_\ell(\bx)/|\bx|^{2\ell+1}$ decaying at infinity have (a distributional)
Fourier transform proportional to $H_\ell(\bk)/\bk^2$ where $H_\ell(\bk)$ is a
homogeneous polynomial of degree $\ell$. Equivalently, when acted on by the
Laplacian this field corresponds to a polynomial source numerator $H_\ell(\bk)$,
as for a linear source. The recursion does not produce such terms beyond linear
order: for $n\ge2$ the convolutions give a non-polynomial dependence on $|\bk|$,
as in the Schwarzschild case where
$J^{(n)}\propto|\bk|^{n-3}$~\cite{Mougiakakos:2024nku}.

For Kerr, dimensional analysis and the symmetry classification allow one
exception, namely  at $n=3$ and spin order $p$ an independent spatial
homogeneous mode with $\ell=p+2$. This mode is not fixed by the nonlinear
Poisson source because it is the freedom to add a decaying solution of the
homogeneous Laplace equation to the particular inverse-Laplacian solution. In
the Fourier representation it has precisely the symmetric trace-free numerator $H_{p+2}(\bk)$
divided by $|\bk|^2$. It must therefore be fixed as boundary or gauge data, rather
than inferred from radial power counting of the nonlinear current. In the
present construction the third-PM boundary prescription sets this addition to
zero. In MPM language it is the order-$\GN^2$ correction to the gauge moment
$X_L$ with even $\ell=p$. The canonical prescription $X_L=0$ makes the same
choice, as verified by the explicit third-PM Kerr metric
in~\eqref{eq:canonical3pm}. A change of the physical multipole moments would
instead change the linear source data, and throughout we keep the Kerr multipole
moments fixed.

\item \emph{Resonances.} The inverse-Laplacian formula~\eqref{eq:invlap} becomes
singular when either factor in its denominator vanishes. This happens precisely
when the would-be particular solution $H_\ell(\bx)/|\bx|^{\mu-2}$ is a solution
of the homogeneous Laplace equation. There are two possibilities: for
$\mu=2\ell+3$, it is the decaying harmonic function
$H_\ell(\bx)/|\bx|^{2\ell+1}$, while for $\mu=2$, it is the growing harmonic
polynomial $H_\ell(\bx)$. In either case, the power-law ansatz is annihilated by
the Laplacian, and a logarithmic term is needed to reproduce the source:
\begin{equation}
\Delta\left(\frac{H_\ell(\bx)\log|\bx|}
{|\bx|^{2\ell+1}}\right)
=-(2\ell+1)\frac{H_\ell(\bx)}{|\bx|^{2\ell+3}},
\quad
\Delta\left(H_\ell(\bx)\log|\bx|\right)
=(2\ell+1)\frac{H_\ell(\bx)}{|\bx|^2}.
\end{equation}
The second resonance is excluded in the present construction. Indeed,
dimensional analysis, since $\hth$ is dimensionless, implies
that a source contribution at PM order \(n\geq2\) and spin order $p$, with STF
degree $\ell$, has radial dependence $H_\ell(\bx)/|\bx|^\mu$ and
$\mu=\ell+n+p+2\geq4$. Thus $\mu=2$ cannot occur.

The first resonance, on the other hand, is possible. Its condition $\mu=2\ell+3$
becomes $\ell=n+p-1$. The possible occurrence of such terms, and their exclusion
in the Kerr recursion considered here, are analysed in
section~\ref{sec:structure}. In dimensional regularisation, the same resonance
manifests itself as a pole at $D=4$ in the inverse-Laplacian
coefficient~\cite{Mougiakakos:2020laz,Mougiakakos:2024nku}.

\end{itemize}

\section{The Kerr source}\label{sec:source}

\subsection{Three-point amplitude and linearised stress tensor}

The Kerr three-point amplitude is the only source datum required for the
linear exterior field in the prescription adopted here. The extended
Polyakov-type hypersurface model of~\cite{BjerrumBohr:2026polyakov} gives an
alternative geometric construction of this same Kerr three-point coupling.
This agreement holds at the level of the one-graviton source and its
multipole moments. It does not establish equivalence of the complete
worldline and hypersurface actions, nor does it fix their possible
higher-multiplicity couplings. The nonlinear exterior problem considered
below is solved using the Einstein equations and the explicit boundary
prescription stated in section~\ref{sec:nonlinwl}.

The classical limit of the three-point amplitude for the emission of a graviton
of momentum $k$ by a massive particle of mass $M$, velocity $v$ and spin tensor
$S^{\mu\nu}$, minimally coupled to gravity, exponentiates in the
spin~\cite{Arkani-Hamed:2017jhn,Guevara:2018wpp,Chung:2018kqs,Arkani-Hamed:2019ymq}.
Stripping the graviton polarisations, it gives the linearised stress tensor of
the Kerr black
hole~\cite{Vines:2017hyw,Guevara:2018wpp,Chung:2019yfs,Gambino:2024uge,Bianchi:2024multipoles}
\begin{equation}\label{eq:Tkerr}
T^{\mu\nu}(k)=2\pi\,\delta(k\cdot v)\left[M\,v^\mu v^\nu\cosh(k\cdot a)+i\,v^{(\mu}S^{\nu)\rho}k_\rho\,\frac{\sinh(k\cdot a)}{k\cdot a}\right],
\end{equation}
where $a^\mu=S^\mu/M$ as before, and the
spin supplementary condition $S^{\mu\nu}v_\nu=0$ holds. The Ward identity $k_\mu
T^{\mu\nu}=0$ follows from $k\cdot v=0$ and the antisymmetry of $S^{\mu\nu}$. In
the rest frame, $v^\mu=(1,\mathbf 0)$, $S^{ij}=\epsilon^{ijk}S^k$ and
$\bk=(k^1,k^2,k^3)$, we have
\begin{equation}\label{eq:Trest}
T^{00}(\bk)=M\cosh(\ba\cdot\bk)\,,\qquad T^{0i}(\bk)=\frac{i}{2}\,(\bk\times\bS)^i\,\frac{\sinh(\ba\cdot\bk)}{\ba\cdot\bk}\,,\qquad T^{ij}(\bk)=0\,,
\end{equation}
with the Fourier convention $T(\bx)=\int\!\frac{d^3k}{(2\pi)^3}\,e^{i\bk\cdot\bx}\,T(\bk)$. The linear coupling~\eqref{eq:Swlexp} of the body, derived in section~\ref{sec:worldline}, and the propagator~\eqref{eq:props} give the first-order field
\begin{equation}\label{eq:J1}
\hth^{(1)\mu\nu}(\bk)=\frac{16\pi\GN}{\bk^2}\,T^{\mu\nu}(\bk)\,.
\end{equation}
For $a=0$ this gives the current
$J^{(1)}_{\mu\nu}=4\rho(|\bk|,4,1)\,\delta^0_\mu\delta^0_\nu$
of~\cite{Mougiakakos:2024nku} used for the Schwarzschild case.

\subsection{Worldline interpretation}\label{sec:worldline}

In position space~\eqref{eq:Trest} reads
\begin{equation}\label{eq:Tx}
T^{00}(\bx)=M\cos(\ba\cdot\nabla)\,\delta^{(3)}(\bx)\,,\qquad
T^{0i}(\bx)=\frac12\left[\nabla\times\Big(\bS\,\frac{\sin(\ba\cdot\nabla)}{\ba\cdot\nabla}\,\delta^{(3)}(\bx)\Big)\right]^i,
\end{equation}
which is the resummation of the spin-induced multipoles of a Kerr black hole,
$\mathcal M_\ell+i\mathcal S_\ell=M(ia)^\ell$~\cite{Hansen:1974zz}. The
covariant version is the exponentiated form
$u^{(\mu}\exp(a*\partial)^{\nu)}{}_\rho u^\rho$ of~\cite{Vines:2017hyw}. The
stress tensor~\eqref{eq:Tx} follows from the linear coupling of the worldline
action of a spinning body with all Wilson coefficients equal to
one~\cite{Levi:2015msa,Vines:2017hyw,Chung:2018kqs}, as we now recall.

\paragraph{Worldline action.} The body couples to the metric only. We
parametrise the worldline $z^\mu(\tau)$ by proper time, $u^\mu=dz^\mu/d\tau$
with $u^2=-1$. The spin is an antisymmetric tensor $S^{\mu\nu}$ with the
covariant supplementary condition $S^{\mu\nu}u_\nu=0$, or equivalently the
vector $S_\mu=-\frac12\epsilon_{\mu\nu\rho\sigma}u^\nu S^{\rho\sigma}$, with
$\epsilon_{0123}=+1$, and we set $a^\mu=S^\mu/M$ the Pauli-Lubanski spin divided by the mass. Up to couplings of higher
order in the curvature, the worldline action of a spinning body
is~\cite{Goldberger:2004jt,Porto:2016pyg,Levi:2015msa,Vines:2017hyw}
\begin{multline}\label{eq:Swl}
S_{\rm wl}=\int d\tau\,\bigg[-M\sqrt{-g_{\mu\nu}u^\mu u^\nu}+\frac12\,S_{\mu\nu}\,
\Omega^{\mu\nu}+M\sum_{n\ge1}\frac{(-1)^{n+1}\,C_{ES^{2n}}}{(2n)!}
\,(a\cdot\nabla)^{2n-2}E_{\mu\nu}\,a^\mu a^\nu\cr
+M\sum_{n\ge1}\frac{(-1)^{n}\,C_{BS^{2n+1}}}{(2n+1)!}\,(a\cdot\nabla)^{2n-1}B_{\mu\nu}\,a^\mu a^\nu\bigg].
\end{multline}
The second term is the kinetic term of the spin with
$\Omega^{\mu\nu}=e^\mu{}_A\,u^\rho \nabla_\rho e^{A\nu}$ being the angular
velocity of a tetrad $e^\mu{}_A$ attached to the body. For a rigid rotator in
flat space $\frac12S_{\mu\nu}\Omega^{\mu\nu}=\bS\cdot\boldsymbol\omega$, so that
$S^{\mu\nu}$ is the angular momentum. Through the Christoffel symbols in
$\nabla_\rho$, this term couples the spin to the gravitational field. The
spin-induced multipoles are built from the electric and magnetic parts of the
curvature,
\begin{equation}\label{eq:EB}
E_{\mu\nu}=R_{\mu\alpha\nu\beta}\,u^\alpha u^\beta\,,\qquad B_{\mu\nu}=\frac12\,\epsilon_{\mu\alpha\rho\sigma}\,R^{\rho\sigma}{}_{\nu\beta}\,u^\alpha u^\beta\,,
\end{equation}
with the Riemann tensor given by 
$R^\rho{}_{\sigma\mu\nu}=\partial_\mu\Gamma^\rho_{\nu\sigma}-\partial_\nu\Gamma^\rho_{\mu\sigma}+\Gamma^\rho_{\mu\lambda}\Gamma^\lambda_{\nu\sigma}-\Gamma^\rho_{\nu\lambda}\Gamma^\lambda_{\mu\sigma}$.
The covariant derivatives act on the curvature only 
\begin{equation}
(a\cdot\nabla)^kE_{\mu\nu}\,a^\mu a^\nu:= a^{\lambda_1}\cdots
a^{\lambda_k}\,a^\mu a^\nu u^\alpha
u^\beta\,\nabla_{\lambda_1}\cdots\nabla_{\lambda_k}R_{\mu\alpha\nu\beta},
\end{equation}
evaluated on the worldline, and similarly for $B$. The point-particle and spin
terms are fixed by general covariance, and the Wilson coefficients $C_{ES^{2n}}$
and $C_{BS^{2n+1}}$ characterise the body. The signs of the operators
in~\eqref{eq:Swl} depend on the conventions for the signature, the curvature and
the anti-symmetric tensor $\epsilon$. We normalised them so that $C_{ES^{2n}}=C_{BS^{2n+1}}=1$ for the
Kerr black hole~\cite{Levi:2015msa,Vines:2017hyw,Chung:2018kqs}, as shown below.
For Kerr the two series resum into
\begin{equation}\label{eq:SKerr}
M\,\frac{1-\cos(a\cdot\nabla)}{(a\cdot\nabla)^2}\,E_{\mu\nu}\,a^\mu a^\nu+M\,\frac{\sin(a\cdot\nabla)-a\cdot\nabla}{(a\cdot\nabla)^2}\,B_{\mu\nu}\,a^\mu a^\nu\,.
\end{equation}
Couplings of higher order in the curvature, such as tidal ones, can be added
to~\eqref{eq:Swl}. They generate nonlinear worldline vertices and are not
included in the exterior problem defined here. Their source-supported effects
and their matching to exterior homogeneous data require additional physical
input, as discussed in section~\ref{sec:nonlinwl}.

\paragraph{Linear coupling.} We expand the worldline action around flat space,
with the body at rest, $z^\mu=\tau v^\mu$, and constant spin. Since $\delta
g_{\mu\nu}:= g_{\mu\nu}-\eta_{\mu\nu}=(P\hth)_{\mu\nu}+\mathcal O(\hth^2)$,
\begin{equation}\label{eq:Swlexp}
S_{\rm wl}=S_{\rm wl}\big|_{g=\eta}+\frac12\int d^Dx\;\hth^{\mu\nu}\Big(T_{\mu\nu}-\frac{1}{D-2}\,\eta_{\mu\nu}T\Big)+\mathcal O(\hth^2)\,,
\end{equation}
where $T^{\mu\nu}=2\,\delta S_{\rm wl}/\delta g_{\mu\nu}|_{g=\eta}$ is the
stress tensor of the body in flat space and $T=\eta_{\mu\nu}T^{\mu\nu}$. In the
rest frame $a\cdot\nabla=\ba\cdot\nabla$, and to linear order we set $\gamma_i:=\delta g_{0i}$. We adopt the rest-frame convention $\epsilon_{0ijk}=+\epsilon_{ijk}$ with $\epsilon_{123}=\epsilon^{123}=+1$. With this orientation and the Riemann-tensor convention specified below~\eqref{eq:EB}, we obtain
\begin{equation}\label{eq:EBlin}
E_{ij}=-\frac12\,\partial_i\partial_j\,\delta g_{00}\,,
\qquad B_{ij}=\frac12\,\partial_j\big(\nabla\times\boldsymbol\gamma\big)_i\,.
\end{equation}
The spin-connection term, with $S^{ik}=\epsilon^{ikm}S^m$, gives
\begin{equation}\label{eq:spinconnectionlin}
\frac12 S^{ik}\Gamma_{k0i}
=\frac12\,\bS\cdot\big(\nabla\times\boldsymbol\gamma\big),
\qquad
\Gamma_{k0i}=\frac12(\partial_i\gamma_k-\partial_k\gamma_i),
\end{equation}
where the positive sign follows from the stated orientation and spin-tensor
convention. This is the linearised contribution of
$\frac12 S_{\mu\nu}\Omega^{\mu\nu}$.
Up to terms with time
derivatives, which are total derivatives along the worldline, we get
\begin{multline}\label{eq:Swllin}
S_{\rm wl}=S_{\rm wl}\big|_{g=\eta}+\frac12\int dt\,
\bigg[M\sum_{n\ge0}\frac{(-1)^n\,C_{ES^{2n}}}{(2n)!}\,(\ba\cdot\nabla)^{2n}\,
\delta g_{00}\cr
+\sum_{n\ge0}\frac{(-1)^n\,C_{BS^{2n+1}}}{(2n+1)!}\,
(\ba\cdot\nabla)^{2n}\;\bS\cdot\big(\nabla\times\boldsymbol\gamma\big)\bigg]_{\bx=0}+\mathcal O(\delta g^2)\,,
\end{multline}
with $C_{ES^0}=C_{BS^1}=1$ for the point-particle and spin terms. The Wilson
coefficients rescale the mass multipole of order $2n$ and the current multipole
of order $2n+1$ with respect to their Kerr values. For Kerr the two sums are
$\cos(\ba\cdot\nabla)$ and $\sin(\ba\cdot\nabla)/(\ba\cdot\nabla)$, and
comparison with $\frac12\int d^Dx\,T^{\mu\nu}\delta g_{\mu\nu}$ gives the stress
tensor~\eqref{eq:Tx}, i.e.\ eq.~\eqref{eq:Tkerr} in momentum space. The terms of
order $\hth^2$ and higher in~\eqref{eq:Swlexp} are the non-linear worldline
couplings. Their treatment in the static point-source prescription is discussed
in section~\ref{sec:nonlinwl}. The body therefore enters only through the source
$j=\frac12PT$, i.e.\ through its three-point amplitude. With the
propagator~\eqref{eq:props} this gives the first-order field~\eqref{eq:J1},
because the trace reversal of the source cancels the numerator of the
propagator.

The trigonometric functions
in~\eqref{eq:Tx} act as differential operators in position space. They become
the hyperbolic form factors of~\eqref{eq:Trest} in momentum space. Since
$\nabla\to i\bk$, the factor $(-1)^n$ in $\mathcal M_{2n}=(-1)^nMa^{2n}$ is
compensated by $i^{2n}$. The two descriptions are therefore identical, and the
hyperbolic functions are not an analytic continuation of the trigonometric ones.

For real $\bk$ the factors $\cosh(\ba\cdot\bk)$ and
$\sinh(\ba\cdot\bk)/(\ba\cdot\bk)$ grow exponentially and are not ordinary
integrable Fourier multipliers. Throughout this paper they are used as
generating functions for the formal multipole expansion in powers of $a$. Each
coefficient is a polynomial in $\bk$ multiplying the $1/\bk^2$ Coulomb
propagator, and its inverse Fourier transform is defined term by term as a
derivative of the Coulomb potential (equivalently, as a distribution away from
the source). This prescription is sufficient to define the exterior field and
does not require assigning an ordinary convergent Fourier integral to the
unexpanded hyperbolic functions. Bounded disk-supported representatives, such as
$\cos(a|\bk_\perp|)$ and $\sin(a|\bk_\perp|)/(a|\bk_\perp|)$, can be used to
encode the same exterior multipole
data~\cite{Israel:1970kp,Bianchi:2024multipoles,Damgaard:2026kerr}. Any equality
between such representatives is understood as equality of their exterior
multipole expansions, up to source-supported terms. We therefore use the
hyperbolic factors only in this formal, multipole-generating sense.

\subsection{The linear field in position space}

The Fourier transforms of the form factors are elementary in flat oblate
spheroidal coordinates $(\varrho,\vartheta,\varphi)$, defined by
\begin{equation}\label{eq:oblate}
x+iy=\sqrt{\varrho^2+a^2}\,\sin\vartheta\,e^{i\varphi}\,,\qquad z=\varrho\cos\vartheta\,,\qquad \Sigma:=\varrho^2+a^2\cos^2\vartheta\,.
\end{equation}
With the coefficientwise multipole prescription described above, for
$\varrho>0$, i.e.\ outside the disk, the exterior transforms are
\begin{equation}\label{eq:FTff}
\begin{aligned}
\int\!\frac{d^3k}{(2\pi)^3}\,e^{i\bk\cdot\bx}\,\frac{4\pi}{\bk^2}\cosh(\ba\cdot\bk)&=\mathrm{Re}\frac{1}{\zeta}=\frac{\varrho}{\Sigma}\,,\\
\int\!\frac{d^3k}{(2\pi)^3}\,e^{i\bk\cdot\bx}\,\frac{4\pi}{\bk^2}\,\frac{\sinh(\ba\cdot\bk)}{\ba\cdot\bk}&=\frac12\int_{-1}^1\frac{d\lambda}{\sqrt{(\bx-i\lambda\ba)^2}}=\frac1a\arctan\frac{a}{\varrho}\,,
\end{aligned}
\end{equation}
where $\zeta:=\sqrt{(\bx-i\ba)^2}=\varrho-ia\cos\vartheta$ is the complex
distance of the Newman--Janis construction~\cite{Newman:1965tw}. Hence
\begin{equation}\label{eq:h1}
\hth^{(1)00}=\frac{4\GN M\varrho}{\Sigma}\,,\qquad
\hth^{(1)0i}=2\GN\Big(\nabla\frac{\arctan(a/\varrho)}{a}\times\bS\Big)^i=\frac{2\GN Ma\,\varrho}{(\varrho^2+a^2)\,\Sigma}\,(\zh\times\bx)^i\,,
\end{equation}
and $\hth^{(1)ij}=0$. These are the $\GN$-linear terms of the Kerr metric in
harmonic coordinates (see section~\ref{sec:chart}). We checked~\eqref{eq:FTff}
and~\eqref{eq:h1} against the multipole series through order $a^{12}$.

\subsection{Non-linear worldline couplings}\label{sec:nonlinwl}

The worldline action~\eqref{eq:Swl} contains, beyond its linear coupling to the
gravitational field, vertices involving two or more gravitons. Such vertices
can contribute to source-supported terms and to the matching of the exterior
solution. 

We define the problem studied here as the stationary exterior
harmonic-gauge Einstein system for $r\neq0$, with the linear Kerr multipoles
fixed by~\eqref{eq:J1} and with the inverse Laplacian prescribed not to add
independent decaying harmonic solutions. This specifies the source data and
the boundary prescription for the exterior recursion. A more complete
worldline effective theory may contain nonlinear source couplings and their
matching can change source-supported terms and, where allowed, the homogeneous
boundary data of the exterior solution. The classification in
section~\ref{sec:structure} identifies the only exceptional homogeneous sector
relevant under the stated symmetry assumptions. Finite-size response
coefficients, including tidal Wilson coefficients, are additional physical
input and must be fixed by matching for a generic compact object.

\section{Residual harmonic-coordinate freedom and all-order uniqueness}\label{sec:structure}

\subsection{Symmetry constraints on the perturbative fields}

At each order $\GN^nM^na^p$, the field generated by the recursion from the
source~\eqref{eq:J1} has the following properties:
\begin{enumerate}
\item[(i)] \emph{Rotational covariance and analyticity.} It is built from $\bx$, $\ba$, $\delta_{ij}$ and $\epsilon_{ijk}$, transforms covariantly under simultaneous rotations of $\bx$ and $\ba$, and is homogeneous of degree $p$ in the components of $\ba$.
\item[(ii)] \emph{Parity.} With $\ba$ treated as an axial vector, the components $\hth^{00}$ and $\hth^{ij}$ are even under $\bx\to-\bx$, whereas $\hth^{0i}$ is odd.
\item[(iii)] \emph{Time reversal.} The stationary solution is invariant under $t\to-t$ together with $\ba\to-\ba$. Thus $\hth^{00}$ and $\hth^{ij}$ are even in $\ba$, and $\hth^{0i}$ is odd in $\ba$.
\end{enumerate}
These properties hold for the source and are preserved by the bilinear
recursion~\eqref{eq:recY}--\eqref{eq:recJ}. Dimensional analysis implies that an
order-$(n,p)$ term is proportional to $|\bx|^{-(n+p)}$ times an angular function
of $\bn=\bx/|\bx|$.

\begin{lemma}\label{lem:degree} Under the conditions (i)--(iii), the order-$(n,p)$
homogeneous tensor structures of $\hth^{00}$ and of the Cartesian components
$\hth^{0i}$ have angular degree at most $p$, while those of $\hth^{ij}$ have
degree at most $p+2$. Only spherical harmonics with $\ell\equiv p\pmod 2$ occur. No
corresponding degree bound for a nonlinear source is assumed here.
\end{lemma}
\begin{proof}
At fixed spin degree $p$, each Cartesian component is a polynomial in the
components of the axial vector $\ba$ and the unit radial vector $\bn$, with
contractions formed using $\delta^{ij}$ and $\epsilon^{ijk}$. A component with
$r$ free spatial indices can therefore have angular degree at most $p+r$.
The time-reversal condition implies that $p$ is even for $\hth^{00}$ and
$\hth^{ij}$, and odd for $\hth^{0i}$. Under spatial parity, the scalar
component is even, the Cartesian vector component is odd, and the rank-two
component is even. Thus the angular degree has the same parity as $p$ in all
three cases. For $\hth^{00}$ this gives an even degree no greater than $p$.
For $\hth^{0i}$ the a priori bound $p+1$ is even and is excluded, leaving a
maximum degree $p$. For $\hth^{ij}$ the maximum allowed degree is $p+2$.
Finally, a polynomial of degree at most $d$ and parity $(-1)^p$ on the sphere
decomposes into spherical harmonics with $\ell=d,d-2,\ldots$, so that
$\ell\equiv p\pmod 2$. These bounds concern the homogeneous tensor structures
of the fields; they do not assert that arbitrary nonlinear products or their
sources obey the same angular-degree bound.
\end{proof}

\subsection{Classification of homogeneous solutions and resonances}

A decaying harmonic function of angular degree $\ell$ has the form
$H_\ell(\bx)/|\bx|^{2\ell+1}$, where $H_\ell(\bx)$ is a homogeneous harmonic
polynomial of degree $\ell$. At order $(n,p)$, dimensional analysis requires
\begin{equation}
\ell=n+p-1.
\end{equation}
For comparison, a logarithmic resonance in the inverse-Laplacian
formula~\eqref{eq:invlap} would arise from a source proportional to
$H_\ell(\bx)/|\bx|^{2\ell+3}$, with the same dimensional relation
$\ell=n+p-1$. The field-structure argument above classifies decaying
homogeneous solutions but it does not, by itself, bound every angular harmonic
in a nonlinear source. We therefore establish the absence of logarithmic
resonances in the actual recursion in the induction proof of
Theorem~\ref{thm:unique}, where the exact Kerr chart supplies,
at each order, a particular solution whose PM coefficient contains no logarithmic term.

\begin{theorem}\label{thm:homog} For $n\geq2$, a decaying homogeneous solution
compatible with (i)--(iii) can occur only in the spatial components $\hth^{ij}$
at $n=3$, with $\ell=p+2$. In particular, no such term is allowed at $n=2$,
where the required degree $\ell=p+1$ has the wrong parity, or at $n\geq4$, where
the required degree exceeds the bound $p+2$.
\end{theorem}

The exceptional sector can be exhibited explicitly by residual harmonic
coordinate transformations. For each even spin order $p=0,2,4,\ldots$, consider
the stationary spatial gauge vector
\begin{equation}\label{eq:exceptionalgaugevector}
\xi^i_{(3,p)}=c_p\, (\GN M)^3(\ba\cdot\nabla)^p\,\partial^i\frac{1}{|\bx|}\,,
\end{equation}
where $c_p$ is an arbitrary dimensionless coefficient. Since
$\Delta(1/|\bx|)=0$ away from the source, each component of $\xi^i_{(3,p)}$ is
harmonic in the exterior, and $\partial_i\xi^i_{(3,p)}=0$. It therefore
preserves the linear harmonic-coordinate condition. The induced linearised
change of the gothic spatial perturbation is, up to the overall sign associated
with the active or passive convention for the coordinate transformation,
\begin{equation}\label{eq:exceptionalgaugemetric}
\delta_\xi\hth^{ij}=
\partial^i\xi^j_{(3,p)}+\partial^j\xi^i_{(3,p)}
-\delta^{ij}\partial_k\xi^k_{(3,p)}
=\partial^i\xi^j_{(3,p)}+\partial^j\xi^i_{(3,p)}.
\end{equation}
It is of order $\GN^3M^3a^p$, falls as $|\bx|^{-(p+3)}$, and its rank-two
angular structure realises the maximal $\ell=p+2$ sector. It has the
rotational, parity and time-reversal properties of the exceptional spatial
sector.

Conversely, this family exhausts the exceptional sector. By dimensional analysis
a term of order $\GN^3M^3a^p$ in the dimensionless field $\hth^{ij}$ is $(\GN
M)^3$ times a function homogeneous of degree $-(p+3)$ in $\bx$. A decaying
homogeneous solution therefore has harmonic components of that degree, which are
combinations of $\partial_L|\bx|^{-1}$ with $|L|=p+2$. By~(i), it is such a
combination contracted with $p$ factors of $\ba$ and with isotropic tensors, and
it carries two free indices. Since $\partial_L|\bx|^{-1}$ is symmetric and
trace-free, each derivative index must be contracted with a factor of $\ba$ or
be a free index, which requires all $p$ factors of $\ba$ and both free indices.
In particular a factor $\ba\cdot\ba$ or a free $\delta^{ij}$ would leave
derivative indices uncontracted, and $\epsilon^{ijk}\partial_k$ is antisymmetric
in $ij$. A Levi-Civita tensor could only enter as $(\ba\times\nabla)^i$, which
leaves a derivative index unmatched. Only
$(\ba\cdot\nabla)^p\partial^i\partial^j|\bx|^{-1}$ remains, which
is~\eqref{eq:exceptionalgaugemetric}. The exceptional sector consequently has
one free coefficient $c_p$ at each even spin order and is generated by a
residual harmonic coordinate transformation. For $p=0$ it is the familiar
one-parameter freedom in the Schwarzschild harmonic radial coordinate discussed
in~\cite{Mougiakakos:2020laz}. There the radial-coordinate shift falls as
$\GN^3M^3/|\bx|^2$, and its gradient generates a spatial metric perturbation
falling as $|\bx|^{-3}$. Symmetry alone therefore does not fix the harmonic
chart in this sector. A boundary prescription is needed to select a
representative.

Time reversal removes one further sector. If the condition~(iii) is dropped,
$\hth^{00}$ is still even in $\ba$, but parity and rotational covariance also
allow $\hth^{0i}$ at even $p$ and $\hth^{ij}$ at odd $p$, with angular degree up
to $p+1$ (from the structures $n^iP_p$ and $n^{(i}(\ba\times\bn)^{j)}P_{p-1}$,
respectively, where $P_k$ is a scalar polynomial of degree $k$ in $\ba$) and
$\ell\equiv p+1\pmod 2$. Decaying homogeneous solutions can then also occur at
$n=2$, with $\ell=p+1$, in $\hth^{0i}$ at even $p$ and in $\hth^{ij}$ at odd
$p$. For $n\geq4$ there are still none. By the argument of
Theorem~\ref{thm:unique} below, a harmonic chart whose PM expansion contains no
logarithmic terms and which obeys (i) and (ii), with linear term~\eqref{eq:h1},
is therefore determined by its homogeneous terms at orders $\GN^2$ and $\GN^3$.
The Cook--Scheel chart is of this type (see section~\ref{sec:CS}).

\subsection{All-order identification under the boundary prescription}

\begin{theorem}\label{thm:unique} Let $\hth_K$ be the expansion of the exact
Kerr gothic metric in a harmonic chart satisfying (i)--(iii), with linear
term~\eqref{eq:h1}, and with no decaying homogeneous contribution in the spatial
components at order $\GN^3$. Consider the stationary, harmonic-gauge Einstein
equations in the exterior, expanded order by order in $\GN$, with the Kerr
multipole source fixed by~\eqref{eq:h1}, the same asymptotic Minkowski
normalisation and mass/current multipoles, and boundary conditions excluding
growing homogeneous modes. The inverse Laplacian is defined without adding
decaying harmonic solutions. The recursion may be represented
by~\eqref{eq:recY}--\eqref{eq:recJ}, or equivalently by the relaxed Einstein
equations of appendix~\ref{app:LL}. Then it is free of logarithmic resonances
and its solution coincides with $\hth_K$ to all orders in $\GN$ and in the spin.
\end{theorem}

\begin{proof}
We proceed by induction in the post-Minkowskian order $n$. Assume that the
recursive field and $\hth_K$ agree through order $n-1$. In the cubic
representation, the auxiliary field at those orders then agrees as well, because
its equation~\eqref{eq:Yeq} is algebraic and is solved order by order. At order
$n$, both fields consequently obey the same Poisson equation~\eqref{eq:recJ}
with the same source.

The PM expansion of the exact Kerr chart contains no logarithmic terms and, by the
induction hypothesis, its lower-order coefficients agree with those of the
recursion. It therefore solves the same order-$n$ Poisson equation with the same
source and provides a particular solution whose PM coefficient contains no
logarithm. A resonant harmonic source cannot be present because it  would force
a logarithm in every solution of that equation. This argument uses the fact that
the exact chart and the recursion implement the same exterior field equations
and the same linear source prescription. The common asymptotic normalisation and
multipole data exclude growing homogeneous solutions, so the difference between
the recursive solution and $\hth_K^{(n)}$ is a decaying homogeneous harmonic
field. By Theorem~\ref{thm:homog}, it vanishes for $n\neq3$. At $n=3$, it is
excluded by the common boundary prescription. The recursion adds no decaying
homogeneous solution, and the chart $\hth_K$ is assumed to contain none. Thus
the fields agree at order $n$, completing the induction.
\end{proof}

The theorem is an identification and uniqueness statement, not an independent
proof of the existence of the Kerr geometry. The exact Kerr solution supplies a
candidate harmonic chart. The theorem shows that the amplitude-based recursion,
with the stated source and boundary prescription, reproduces that chart. In
particular, the symmetry argument by itself does not remove the exceptional
$n=3$ freedom. Its removal uses the explicit boundary condition on that sector.

\section{The harmonic coordinates selected by the recursion}\label{sec:chart}

\subsection{Harmonic functions on the Kerr background}

In Boyer--Lindquist coordinates
$(t,r,\theta,\phi)$~\cite{Kerr:1963ud,Boyer:1966qh}, setting $R(r):=r-G_NM$ 
and $\Delta(r):=R(r)^2+a^2-\GN^2 M^2$, a stationary function $\Psi(r,\theta,\phi)$
is harmonic if and only if
\begin{equation}\label{eq:waveq}
\partial_r\big(\Delta(r)\,\partial_r\Psi(r,\theta,\phi)\big)
+\frac{1}{\sin\theta}\partial_\theta\big(\sin\theta\,\partial_\theta\Psi(r,\theta,\phi)\big)
+\Big(\frac{1}{\sin^2\theta}-\frac{a^2}{\Delta(r)}\Big)\partial_\phi^2\Psi(r,\theta,\phi)=0\,.
\end{equation}
The equation separates on $P_\ell^m(\cos\theta)e^{im\phi}$. Harmonic Cartesian
coordinates behave like $\ell=1$ functions at infinity. For
$\Psi(r,\theta,\phi)=H(r)\cos\theta$ and $\Psi(r,\theta,\phi)=F(r)\sin\theta\,e^{i\phi}$,  respectively, the radial equations
are
\begin{equation}\label{eq:radial}
\partial_r\big(\Delta(r) \partial_r H(r)\big)-2H(r)=0\,,\qquad \partial_r\big(\Delta(r) \partial_rF(r)\big)+\left(\frac{a^2}{\Delta(r)}-2\right)\,F(r)=0\,.
\end{equation}
The solutions growing at infinity are $H(r)=R(r)$ and
\begin{equation}\label{eq:Fpm}
F_\pm(r)=(R(r)\pm ia)\,e^{\mp i\Theta(r)}\,,\qquad \Theta'(r)=-\frac{a}{\Delta(r)}\,,
\end{equation}
with $\Theta(r)$ as in~\eqref{eq:Theta}. Indeed, 
one checks~\eqref{eq:radial} for any $\Theta(r)$ with
$\partial_r\Theta(r)=-a/\Delta(r)$. The radial equation is real, so
$F_-(r)=\overline{F_+}(r)$ and $\mathrm{Re}\,F_+(r)$ are solutions too.

\subsection{Three harmonic charts}\label{sec:threecharts}

The spatial harmonic coordinates of Cook and Scheel~\cite{Cook:1997qc}, together
with the Boyer--Lindquist time, define the chart
\begin{equation}\label{eq:CS}
\text{CS:}\qquad T=t\,,\qquad x+iy=(R(r)+ia)\,e^{i(\phi-\Theta(r))}\sin\theta\,,
\qquad z=R(r)\cos\theta\,.
\end{equation}
This is the chart of the closed-form harmonic Kerr metric of~\cite{Lin:2014laa},
with which the recursion was compared in~\cite[eqs.~(2.5)--(2.6)]{Damgaard:2026kerr}
 (see also~\cite{Xing:2026rrd}). Since
$d(\phi-\Theta(r))=d\phi+a\,dr/\Delta(r)$, its azimuthal coordinate is the ingoing
Kerr azimuth. The choice $T=t$ is nevertheless the Boyer--Lindquist time, so the
displayed four-coordinate chart should not be interpreted as a
horizon-penetrating chart. The chart built on $F_-(r)$ is the corresponding
time-reversed radial representative, with its azimuthal coordinate adapted to the
outgoing Kerr azimuth. Under time reversal, combined with $\ba\to-\ba$ to map
the Kerr metric back onto itself, the two radial representatives are exchanged.
Neither is time-reversal covariant. The covariant combination is their average,
which is the chart~\eqref{eq:mainchart} 
\begin{equation}\label{eq:chart}
T=t\,,\qquad x+iy=\sqrt{R(r)^2+a^2}\,\cos\delta(r)\,\sin\theta\,e^{i\phi}\,,
\qquad z=R(r)\cos\theta\,,
\end{equation}
where
\begin{equation}\label{eq:delta}
\begin{aligned}
\delta(r)=\Theta(r)-\arctan\frac{a}{R(r)}&=\int_R^\infty\frac{a\,\GN^2M^2\,du}{(u^2+a^2)(u^2+a^2-\GN^2M^2)}\\
&=\sum_{k\ge1}(\GN M)^{2k}\int_R^\infty\frac{a\,du}{(u^2+a^2)^{k+1}}\,.
\end{aligned}
\end{equation}
In particular
\begin{equation}\label{eq:deltasmall}
\delta(r)=\frac{\GN^2M^2}{2a^2}\left(\arctan\frac{a}{R(r)}-\frac{aR(r)}{R(r)^2+a^2}\right)+\mathcal O(\GN^4)
=\frac{\GN^2M^2a}{3R(r)^3}+\cdots.
\end{equation}
In the chart~\eqref{eq:chart} the azimuth is the Boyer--Lindquist azimuth, and
the deviation from flat oblate spheroidal coordinates (with $\varrho=R(r)$) is the
factor $\cos\delta=1+\mathcal O(\GN^4)$. The CS chart is obtained from~\eqref{eq:chart}
by the rotation $\varphi\to\varphi-\delta(r)$ about the spin axis, together with
a rescaling of the cylindrical radius by $1/\cos\delta$.

\subsection{Exact gothic metric}

We express the gothic metric of~\eqref{eq:chart} in the orthonormal cylindrical
frame $(\partial_T,\evarpi,\ephi,\zh)$ attached to the harmonic point $\bx$,
with $\varpi=|\bx_\perp|$. Define
\begin{equation}
F(r)=\sqrt{R(r)^2+a^2}\,\cos\delta(r)\,,
\end{equation}
so that
\begin{equation}
    \begin{aligned}
\partial_r F(r)&=\frac{R(r)\cos\delta(r)}{\sqrt{R(r)^2+a^2}}-\sqrt{R(r)^2+a^2}\,\partial_r\delta(r)\sin\delta(r)\,,\cr
\partial_r\delta(r)&=-\frac{a\,\GN^2M^2}{(R(r)^2+a^2)\Delta(r)}\,,
    \end{aligned}
\end{equation}
and setting 
\begin{equation}
\Sigma_R:= R(r)^2+a^2\cos^2\theta\,,
\end{equation}
we have
\begin{equation}
\cD=\cos^2\!\delta(r)\;\Sigma_R+\frac{a\GN^2M^2R(r)}{\Delta(r)}\,\sin^2\theta\,\sin\delta(r)\cos\delta(r)\,.
\end{equation}
Then the non-vanishing inverse gothic metric components are
\begin{align}
\gth^{TT}&=-\frac{(r^2+a^2)^2-a^2\Delta(r)\sin^2\theta}{\Delta(r)\,\cD}\,,\qquad
\gth^{T\hat\varphi}=-\frac{2\GN Ma\,r\,F(r)\sin\theta}{\Delta(r)\,\cD}\,,\label{eq:gexact1}\\
\gth^{\hat\varpi\hat\varpi}&=\frac{F'(r)^2\Delta(r)\sin^2\theta+F(r)^2\cos^2\theta}{\cD}\,,\qquad
\gth^{\hat\varphi\hat\varphi}=\frac{F(r)^2\,(\Delta(r)-a^2\sin^2\theta)}{\Delta(r)\,\cD}\,,\label{eq:gexact2}\\
\gth^{\hat z\hat z}&=\frac{\Delta(r)\cos^2\theta+R(r)^2\sin^2\theta}{\cD}\,,\qquad
\gth^{\hat\varpi\hat z}=\frac{\sin\theta\cos\theta\,(F'(r)\Delta(r)-R(r)F(r))}{\cD}\,.\label{eq:gexact3}
\end{align}
Equivalently $\hth^{0i}=W\,(\zh\times\bx)^i$ with $W=2\GN Ma\,r/(\Delta(r)\cD)$. In
the CS chart~\eqref{eq:CS} the components~\eqref{eq:gexact1}--\eqref{eq:gexact3}
hold with $\delta=0$ in $F(r)$, $\partial_r F(r)$ and $\cD$, except for
\begin{equation}\label{eq:CSframe}
\gth^{\hat\varphi\hat\varphi}_{\rm CS}=1-\frac{a^2\GN^2M^2\sin^2\theta}{(R(r)^2+a^2)\Sigma_R}\,,\qquad
\gth^{\hat\varpi\hat\varphi}_{\rm CS}=\frac{a\GN^2M^2R(r)\sin^2\theta}{(R(r)^2+a^2)\Sigma_R}\,,\qquad
\gth^{\hat\varphi\hat z}_{\rm CS}=\frac{a\GN^2M^2\sin\theta\cos\theta}{\sqrt{R(r)^2+a^2}\,\Sigma_R}\,,
\end{equation}
so that all its spatial components are polynomials of degree two in $\GN M$
in agreement with~\cite{Damgaard:2026kerr}. For $a=0$,
eqs.~\eqref{eq:gexact1}--\eqref{eq:gexact3} give the components of the Schwarzschild metric
$\gth^{TT}=-(1+\rho)^3/(1-\rho)$ and $\gth^{ij}=\delta^{ij}-\rho^2n^in^j$ with
$\rho=\GN M/R$ the expansion parameter introduced in~\cite{Mougiakakos:2024nku}.

\paragraph{Coordinate Jacobian and gothic density.}
The coordinate map is particularly simple in cylindrical variables. Using
$R=r-\GN M$ as the radial variable, define
\begin{equation}
\widehat F(R):=F(r(R))
=\sqrt{R^2+a^2}\cos\delta(R), \qquad r=R+\GN M.
\end{equation}
As a consistency check we notice that $\widehat F(R)\to R$ in the Schwarzschild
limit $a\to0$ and $R>0$ using $\delta(R)\to0$ from~\eqref{eq:delta}. 

The harmonic cylindrical radius and the Cartesian harmonic coordinate $z$ are then
\begin{equation}
\varpi=\widehat F(R)\sin\theta,
\qquad z=R\cos\theta.
\end{equation}
The nontrivial meridional Jacobian is
\begin{equation}
\mathcal J
=\det\frac{\partial(\varpi,z)}{\partial(R,\theta)}
=-R\partial_R\widehat F(R)\sin^2\theta
-\widehat F(R)\cos^2\theta.
\label{eq:meridionalJacobian}
\end{equation}
The Cartesian harmonic coordinates are $x=\varpi\cos\varphi$,
$y=\varpi\sin\varphi$, and $z=R\cos\theta$, with $T=t$ and $\varphi$ unchanged.
Thus the full spatial Jacobian from $(R,\theta,\varphi)$ to $(x,y,z)$ has
determinant $\varpi\mathcal J$. Starting from the Boyer--Lindquist line element,
the covariant metric in $X^A=(T,x,y,z)$ is obtained by the tensorial pullback
\begin{equation}
g_{AB}=\frac{\partial x^\mu_{\rm BL}}{\partial X^A}
\frac{\partial x^\nu_{\rm BL}}{\partial X^B}g^{\rm BL}_{\mu\nu}.
\label{eq:metricpullback}
\end{equation}
The gothic inverse metric is then formed as a Cartesian tensor density,
\begin{equation}
\mathfrak g^{AB}=\sqrt{-g_{(X)}}\,g^{AB}_{(X)},\qquad
\sqrt{-g_{(X)}}=\frac{\sqrt{-g_{\rm BL}}}{\varpi|\mathcal J|}.
\label{eq:gothicdensity}
\end{equation}
Finally, a rotation of the Cartesian spatial indices to the local orthonormal
cylindrical basis $(\evarpi,\ephi,\zh)$ gives the components displayed
in~\eqref{eq:gexact1}--\eqref{eq:gexact3}. This procedure fixes both the density
weight and the off-diagonal component conventions. The map from the
Boyer--Lindquist variables $(R,\theta,\varphi)$ to the Cartesian harmonic
coordinates $(x,y,z)$ is regular away from the axis, provided $\widehat
F(R)\neq0$ and $\mathcal J\neq0$. If $\widehat F(R)=0$, the cylindrical radius
$\varpi=\widehat F(R)\sin\theta$ vanishes for every $\theta$, so that the
azimuthal circles collapse and the spatial Jacobian $\varpi\mathcal J$
degenerates.  For $|a|<\GN M$, $\Theta$ diverges logarithmically at the
horizon $R=q$, so $\widehat F(R)$ first vanishes at a radius $R_*>q$ where
$\delta(R_*)=\pi/2$, with
$R_*-q\simeq2q\exp\!\big[-2q\big(\tfrac\pi2+\arctan\tfrac aq\big)/a\big]$ for
small spin. The expressions above are understood on the asymptotic coordinate
patch, away from such degeneracies.

\subsection{The recursion produces the time-symmetric chart}

\begin{theorem}\label{thm:chart} The chart~\eqref{eq:chart} satisfies the
hypotheses of Theorem~\ref{thm:unique}, including the boundary prescription in
the exceptional spatial $n=3$ sector. Consequently, the harmonic-gauge Einstein
recursion with the Kerr multipole source~\eqref{eq:J1} reproduces, to all orders
in $\GN$ and in the spin, the Kerr metric in the harmonic
coordinates~\eqref{eq:mainchart}. The cubic
recursion~\eqref{eq:recY}--\eqref{eq:recJ} is an efficient representation of
this system.
\end{theorem}
\begin{proof}
The coordinates are harmonic by~\eqref{eq:radial}, and $T=t$ is harmonic by
stationarity and axisymmetry. Time-reversal covariance follows because the
selected radial function is $\operatorname{Re}F_+$ and is even under $a\to-a$
when the spin reversal is included in the time-reversal map.
Equation~\eqref{eq:delta} shows that $\delta=\mathcal O(\GN^2)$ and is
analytic in the spin near $a=0$. Together with the flat oblate-coordinate map,
this gives the analyticity and covariance properties (i) and (ii) away from
the source. To make explicit which spatial variables are
held fixed in the PM expansion, we use the flat oblate spheroidal coordinates
$(\varrho,\vartheta,\varphi)$ of the harmonic Cartesian point,
eq.~\eqref{eq:oblate}, for which $\varpi=\sqrt{\varrho^2+a^2}\sin\vartheta$,
$z=\varrho\cos\vartheta$ and $\varphi=\phi$.
The exact chart~\eqref{eq:chart} gives instead
$\varpi=\widehat F(R)\sin\theta$ and $z=R\cos\theta$.
Writing $\varrho=R+u\,\delta^2+\mathcal O(\delta^4)$ and
$\vartheta=\theta+v\,\delta^2+\mathcal O(\delta^4)$, inversion of these two
relations yields
\begin{equation}\label{eq:oblateinversion}
\begin{aligned}
\varrho&=R-\frac{R(R^2+a^2)\sin^2\theta}{2\Sigma_R}\,\delta^2
+\mathcal O(\delta^4),\\
\vartheta&=\theta-\frac{(R^2+a^2)\sin\theta\cos\theta}{2\Sigma_R}\,\delta^2
+\mathcal O(\delta^4),\\
\varphi&=\phi,
\end{aligned}
\qquad \Sigma_R=R^2+a^2\cos^2\theta.
\end{equation}
Since $\delta=\mathcal O(\GN^2)$, this proves
$(\varrho,\vartheta,\varphi)=(R,\theta,\phi)+\mathcal O(\GN^4)$. Thus the PM
expansion through third order may equivalently be performed at fixed flat oblate
coordinates or at fixed $(R,\theta,\phi)$. The coordinates first differ at order
$\GN^4$, which affects the metric at fixed $\bx$ from order $\GN^5$ (see
section~\ref{sec:PM}). The linear term is~\eqref{eq:h1}. At fixed $(R,\theta)$
the spatial components~\eqref{eq:gexact2}--\eqref{eq:gexact3} contain $M$ only
through $\Delta$ and $\delta$, that is only in the combination $\GN^2M^2$
by~\eqref{eq:delta}. The same holds for the relation~\eqref{eq:chart} between
$(R,\theta)$ and the harmonic point $\bx$, which contains $M$ only through
$\delta$. The spatial components therefore contain only even powers of $\GN$ at
fixed $\bx$, and in particular no term of order $\GN^3$. Neither the inversion
nor the change to Cartesian components introduces odd powers of $\GN$.
By~\eqref{eq:delta}, $\delta$ is a series in $\GN^2M^2$ with coefficients
depending on $R$ and $a$ only, so the inverse map $R(\bx),\theta(\bx)$ is a
series in $\GN^2$, as~\eqref{eq:oblateinversion} shows at first order. The
Cartesian components follow from the cylindrical frame components by a rotation
through the azimuth $\phi$, which does not involve $M$. The variable $R$ matters
here, since the Boyer--Lindquist radius differs from it by a term linear in $\GN
M$. This fixes the exceptional homogeneous sector required by
Theorem~\ref{thm:unique}.
\end{proof}

\subsection{Relation to the Cook--Scheel chart}\label{sec:CS}

The relation between $(R,\theta,\phi)$ and the flat oblate coordinates
$(\varrho,\vartheta,\varphi)$ is given explicitly in
\eqref{eq:oblateinversion}. The two radial and polar coordinate
pairs agree through third PM order. At order $\GN^2$ the CS chart differs from~\eqref{eq:chart} by the linearised gauge transformation generated by the rotation $\varphi\to\varphi-\delta(r)$:
\begin{equation}\label{eq:CSdiff}
\hth^{ij}_{\rm CS}-\hth^{ij}=-\frac{\GN^2M^2a\,\sin\vartheta}{\sqrt{\varrho^2+a^2}\;\Sigma}\left(\partial_\varrho x^i\,\hat e_\varphi^j+\hat e_\varphi^i\,\partial_\varrho x^j\right)+\mathcal O(\GN^4)\,,
\end{equation}
whose leading term for small spin is
$-2\GN^2M^2a\,x^{(i}(\zh\times\bx)^{j)}/|\bx|^5$. 
To derive~\eqref{eq:CSdiff}, note that $\cos\delta=1+\mathcal O(\GN^4)$, so at
this order the CS coordinates follow from~\eqref{eq:chart} by the shift
$\zeta^i=-\delta(\varrho)\,(\zh\times\bx)^i$, and
$\hth^{ij}_{\rm CS}-\hth^{ij}=-(\partial^i\zeta^j+\partial^j\zeta^i-\delta^{ij}\partial_k\zeta^k)$.
Since $\partial_k\zeta^k=0$ and the derivative of $\zh\times\bx$ is
antisymmetric, only $\delta'(\varrho)=-\GN^2M^2a/(\varrho^2+a^2)^2+\mathcal O(\GN^4)$
contributes, therefore the derivative removes the $\arctan(a/\varrho)$ of $\delta_2$.
With $\nabla\varrho=(\varrho^2+a^2)\,\partial_\varrho\bx/\Sigma$ and
$\zh\times\bx=\sqrt{\varrho^2+a^2}\sin\vartheta\,\hat e_\varphi$ this
gives~\eqref{eq:CSdiff}, whose overall sign is that of $\delta'$.

The term in~\eqref{eq:CSdiff} is odd in $a$ and harmonic. It is the term
identified at order $\GN^2$ in~\cite{Damgaard:2026star,Damgaard:2026kerr}. The
residual gauge vector given there to all orders in
$a$~\cite[eq.~(52)]{Damgaard:2026star} and~\cite[eq.~(2.21)]{Damgaard:2026kerr}
generates the rotation by the order-$\GN^2$ term of $\delta$
in~\eqref{eq:delta}. Theorem~\ref{thm:homog} explains why the recursion cannot
produce it. Time reversal forbids homogeneous terms at order $\GN^2$. The term
originates in the ingoing azimuth of the CS chart, which breaks time reversal
(see section~\ref{sec:chart}), not in the recursion. The CS gothic metric is
exactly quadratic in $\GN$ in its spatial components, while that
of~\eqref{eq:chart} contains all even orders in $\GN$. Matching the recursion
against the CS chart therefore fails at order $\GN^2a$.

Conversely, the CS chart obeys (i) and (ii) and has the linear
term~\eqref{eq:h1}. It shares the time coordinate of~\eqref{eq:chart}, so that
at order $\GN^2$ the two metrics differ only by~\eqref{eq:CSdiff}, and its
spatial components have no term of order $\GN^3$. By the remark following
Theorem~\ref{thm:homog}, the CS metric is therefore the solution of the
recursion in which the order-$\GN^2$ term of~\eqref{eq:CSdiff} is added, and no
other homogeneous term. This is the procedure of~\cite{Damgaard:2026kerr}, where
agreement with~\cite{Lin:2014laa} was found through fourth order. The argument,
presented in the present paper, shows that the agreement holds to all orders. It
also shows that, within the conditions (i)--(iii), the only possible residual
homogeneous terms occur at orders $\GN^2$ and $\GN^3$ but time reversal excludes
the former, while the boundary prescription fixes the latter. The higher-order
terms of the rotation by $\delta$ and the rescaling of the cylindrical radius by
$1/\cos\delta$ are then generated by the recursion itself. Physical observables
are of course insensitive to the choice of chart.

Without the gauge term, the recursion of~\cite{Damgaard:2026kerr} solves the
same harmonic field equations with the same linear field and obeys (i)--(iii).
At third order it reproduces the expansion of the metric of~\cite{Lin:2014laa}
given in~\cite[eqs.~(C.3)--(C.4)]{Damgaard:2026kerr}, which coincides with that
of~\eqref{eq:chart} at this order. The order-$\GN^2$ rotation leaves the
axisymmetric linear field invariant, and the other differences between the two
charts are of order $\GN^4$. This is consistent with the observation
of~\cite{Damgaard:2026kerr} that the gauge term affects the recursion only from
fourth order on. Through third order the gauge-term-free result therefore
coincides with~\eqref{eq:chart}, and by the argument of Theorem~\ref{thm:unique}
it does so to all orders. We checked this directly at fourth order: the
components given in~\cite[eqs.~(5.19), (D.19), (D.20)]{Damgaard:2026kerr} as
finite sums of Jacobi polynomials, for each power of $a$, agree
with~\eqref{eq:h004}--\eqref{eq:hij4} (see section~\ref{sec:checks}), where
these sums are resummed by $\delta_2$, which contains
$\arctan(a/\varrho)$.\footnote{In the fourth-PM notation of~\cite[eq.~(D.20)]{Damgaard:2026kerr} we read
$h'^{12}$ as the coefficient of $a^n$, $n$ even. In $h'^{11}$ we take $m=n-1$
and the prefactor $M^4/r^{2n+6}$, and include in the coefficient of $x^2$ the
sum that appears in $h'^{12}$, as required by axial symmetry, $h'^{11}-h'^{22}=(x^2-y^2)\,h'^{12}/(xy)$.} The
gauge-term-free result is therefore not rational, in line with the observation
of~\cite{Damgaard:2026kerr}, but it has the closed
form~\eqref{eq:gexact1}--\eqref{eq:gexact3}. Redundant terms are therefore not
needed to re-sum the series into a closed form, only to reach the particular
rational form of~\cite{Lin:2014laa}.

\subsection{Relation to canonical harmonic coordinates}\label{sec:canonical}

The general harmonic MPM solution is parametrised by the mass and current
moments and by four families of gauge moments, $(W_L,X_L,Y_L,Z_L)$, all of
which may be power series in $\GN$. The canonical representative is the one in
which the gauge moments vanish at every
order~\cite{Blanchet:1985sp,Blanchet:2013haa}. This prescription fixes a
representative within the MPM construction and it should not be confused with a
general theorem that all harmonic charts coincide. The MPM variable is
$h^{\mu\nu}:=\gth^{\mu\nu}-\eta^{\mu\nu}=-\hth^{\mu\nu}$. The gauge moments
enter the linear field through a gauge vector, and for a stationary field the
term of order $\GN^n$, with $n\geq2$, is constructed as
\begin{equation}\label{eq:MPMalgorithm}
h^{\mu\nu}_{(n)}=u^{\mu\nu}_{(n)}+v^{\mu\nu}_{(n)}\,,\qquad
u^{\mu\nu}_{(n)}=\mathop{\mathrm{FP}}\Delta^{-1}\Lambda^{\mu\nu}_{(n)}\,,\qquad
v^{\mu\nu}_{(n)}=\mathcal V^{\mu\nu}\big[\partial_\rho u^{\rho\sigma}_{(n)}\big]\,.
\end{equation}
Here $\Delta h_{(n)}=\Lambda_{(n)}$ is the term of order $\GN^n$ of the relaxed
equations~\eqref{eq:LLwave} written for $h^{\mu\nu}$, so that $\Lambda_{(n)}$
is built from the lower-order fields, $\mathrm{FP}$ denotes the finite part of
the regularised inverse Laplacian, and $\mathcal V$ is a linear map that
restores the harmonic condition. On a non-resonant term
$H_\ell(\bx)/|\bx|^{\mu}$ of the source, with $\mu\neq2\ell+3$, no finite part
is needed and $\mathrm{FP}\,\Delta^{-1}$ acts
as~\eqref{eq:invlap}~\cite[eq.~(17)]{Xing:2026rrd}.

For Kerr, Xing, He and Cao~\cite{Xing:2026rrd} determine the canonical moments
from the Geroch--Hansen multipoles and Gürsel's theorem, construct the
canonical metric explicitly through fourth PM order, to all orders in the
spin, and verify the vacuum Einstein equations to that order. Denoting by
$h_{\mathrm{can}(n)}$ the term of order $\GN^n$ of the canonical metric, their
closed forms for $n\leq4$ are minus~\eqref{eq:h1},
\eqref{eq:h00PM}--\eqref{eq:hijPM} and~\eqref{eq:h004}--\eqref{eq:hij4}, for
all components and exactly in the spin. At third order, for instance,
\begin{equation}\label{eq:canonical3pm}
 h^{00}_{\mathrm{can}(3)}=-\frac{8\GN^3M^3\varrho}{(\varrho^2+a^2)\,\Sigma}\,,\qquad
 h^{ij}_{\mathrm{can}(3)}=0\,,\qquad
 h^{0i}_{\mathrm{can}(3)}=-\frac{2\GN^3M^3a\varrho}{(\varrho^2+a^2)^2\,\Sigma}\,
 (\zh\times\bx)^i\,,
\end{equation}
in the flat oblate coordinates~\eqref{eq:oblate} of the harmonic point. This is
minus the third-order term of~\eqref{eq:h00PM}--\eqref{eq:hijPM}, including the
vanishing of the spatial components in the exceptional sector.

The homogeneous solutions classified in section~\ref{sec:structure} are, in
MPM language, higher-PM corrections to the gauge moments. The
gauge vector~\eqref{eq:exceptionalgaugevector} is a pure gradient, which is the
form in which the moments $X_L$ enter the linear gauge vector. The coefficient
$c_p$ of the exceptional sector is therefore a correction of order $\GN^2$ to
$X_L$, with $\ell=p$ even. Similarly, the analysis of the Jiang--Lin
metric~\cite{Lin:2014laa} in~\cite{Xing:2026rrd} finds that at first order in
$\GN$ its only nonzero gauge moments are the $Z_L$ with $\ell$ odd. The terms
that they generate in the metric at order $\GN^2$ are those
of~\eqref{eq:CSdiff}, with $h=-\hth$. These gauge moments are thus the
time-odd homogeneous term of section~\ref{sec:CS}, which enters the metric at
second PM order, and they are distinct from the exceptional third-order
sector. By
Theorem~\ref{thm:homog} and the discussion following it, the 
order-$\GN^2$ corrections to $X_L$ are the only 
higher-order gauge-moment freedom compatible with (i)--(iii). 
The canonical prescription sets these corrections to zero.

\begin{theorem}\label{thm:canonical} The canonical MPM metric with the Kerr mass
and current moments, with all four families of gauge moments $(W_L,X_L,Y_L,Z_L)$
set to zero at every PM order and with the inverse-Laplacian
prescription~\eqref{eq:MPMalgorithm}, is the Kerr metric in the
chart~\eqref{eq:chart}: $h^{\mu\nu}_{\mathrm{can}(n)}=-\hth^{(n)\mu\nu}$ at
every PM order $n$ and to all orders in the spin. In
particular~\eqref{eq:mainchart} is the transformation from Boyer--Lindquist to
canonical harmonic coordinates.
\end{theorem}
\begin{proof}
The canonical linear metric with the Kerr moments is minus~\eqref{eq:h1}.
Assume that the canonical metric and the exact-chart metric agree below order
$n$, so that $h^{\mu\nu}_{\mathrm{can}(m)}=-\hth^{(m)\mu\nu}$ for $m<n$.
Their order-$n$ relaxed equations then have the same source
$\Lambda^{\mu\nu}_{(n)}$
\begin{equation}
\Delta h^{\mu\nu}_{\mathrm{can}(n)}=\Lambda^{\mu\nu}_{(n)},
\qquad
\Delta\big(-\hth^{(n)\mu\nu}\big)=\Lambda^{\mu\nu}_{(n)}.
\label{eq:MPMsource}
\end{equation}
The exact chart supplies a solution without logarithmic terms. A resonant
source would require a logarithm in every particular solution, so no resonant
term occurs at this order and the finite-part operation is unnecessary.
The source $\Lambda^{\mu\nu}_{(n)}$ is constructed from the lower-order
fields by the relaxed Einstein equations. By induction, these fields
have the rotational covariance, parity and time-reversal properties
listed in section~\ref{sec:structure}. The ordinary inverse Laplacian
commutes with rotations and parity and preserves the combined
time-reversal transformation $t\to-t$, $\ba\to-\ba$. It also preserves
the expansion in homogeneous powers of the spin. Consequently, the
MPM particular solution $u^{\mu\nu}_{(n)}$ obeys the same symmetry
properties as the exact-chart coefficient $-\hth^{(n)\mu\nu}$.

It remains to identify the MPM particular solution $u_{(n)}$ with the exact
field. This is the step for which the homogeneous-mode classification is
needed. Define
\begin{equation}
 w^{\mu\nu}_{(n)}:=u^{\mu\nu}_{(n)}+\hth^{(n)\mu\nu}.
 \label{eq:MPMhomdiff}
\end{equation}
Since $u_{(n)}$ and $-\hth^{(n)}$ solve the same Poisson equation, $\Delta
w^{\mu\nu}_{(n)}=0$. The common asymptotic normalisation excludes non-decaying
homogeneous solutions, while asymptotic flatness excludes growing ones. Fixing
the mass and current multipoles fixes the physical multipole data. By the
symmetry argument above, the difference $w^{\mu\nu}_{(n)}$ satisfies the
hypotheses of Theorem~\ref{thm:homog}. It follows that $w_{(n)}=0$ for $n\ne3$.

At $n=3$, the only possible difference is the exceptional spatial mode
identified in Section~4. By Theorem~\ref{thm:chart}, $\hth^{(3)ij}=0$
identically in the exact chart, so Eq.~\eqref{eq:MPMsource} gives
$\Lambda^{(3)ij}=\Delta(-\hth^{(3)ij})=0$. Since the canonical inverse-Laplacian
operator $\mathrm{FP}\,\Delta^{-1}$ in Eq.~\eqref{eq:invlap} is linear and adds
no homogeneous mode in the canonical prescription, it follows from
$\Lambda^{(3)ij}=0$ that $u^{(3)ij}=0$, and hence $w^{(3)}=0$.

We have therefore established $u^{\mu\nu}_{(n)}=-\hth^{(n)\mu\nu}$ at every PM
order. Since the exact chart is harmonic,
$\partial_\rho u^{\rho\sigma}_{(n)}=0$. The gauge-restoring contribution
defined in~\eqref{eq:MPMalgorithm} is a linear functional of this divergence
and hence vanishes, $v^{\mu\nu}_{(n)}=0$. It follows that
\begin{equation}
h^{\mu\nu}_{\mathrm{can}(n)}
=u^{\mu\nu}_{(n)}+v^{\mu\nu}_{(n)}
=-\hth^{(n)\mu\nu},
\end{equation}
which completes the induction.
\end{proof}

Theorem~\ref{thm:canonical} has the following consequences for the canonical
Kerr metric, at every PM order:
\begin{enumerate}
\item[(a)] no logarithm arises and no finite part is needed in the
iteration~\eqref{eq:MPMalgorithm};
\item[(b)] the homogeneous term $v_{(n)}$ vanishes;
\item[(c)] the components $h^{00}_{\mathrm{can}}$ and $h^{ij}_{\mathrm{can}}$
are even in $a$, and $h^{0i}_{\mathrm{can}}$ is odd;
\item[(d)] the spatial components contain only even powers of $\GN$, so that
$h^{ij}_{\mathrm{can}(n)}=0$ for odd $n$.
\end{enumerate}
Properties (a) and (b) are part of the proof, (c) is the time-reversal
covariance of the chart, and (d) is shown in the proof of
Theorem~\ref{thm:chart}. In~\cite{Xing:2026rrd}, properties (a) and (b) were
verified through fifth PM order and adopted as assumptions beyond that order.
Property (c), the even or odd spin-parity structure, was observed there through
fourth PM order and supported by an induction argument that relies on the same
assumption. Property (d), the vanishing of the spatial components at odd PM
orders, was checked through fifth PM order but was not proved to all orders. The
present construction establishes these properties to all PM orders. As an
additional computational check of (a) and (b), we carried out
the iteration~\eqref{eq:MPMalgorithm} in exact arithmetic through order $\GN^8$
and spin order $a^{12}$: no resonant term occurs, $\partial_\mu
u^{\mu\nu}_{(n)}$ vanishes at each order, and the iteration reproduces the
output of the cubic recursion (see section~\ref{sec:checks}).

The theorem also gives the coordinate transformations explicitly.  The
transformation from Boyer--Lindquist to canonical harmonic coordinates is
obtained in~\cite[eqs.~(66)--(69)]{Xing:2026rrd} through order $\GN^3$. At this
order it reduces to $x+iy=\sqrt{R^2+a^2}\,\sin\theta\,e^{i\phi}$ and
$z=R\cos\theta$ with $R=r-\GN M$, in agreement with~\eqref{eq:chart} since
$\cos\delta=1+\mathcal O(\GN^4)$. The transformation at fourth order is left
open there. It is given by~\eqref{eq:oblateinversion}, and to all orders by the
factor $\cos\delta$ in~\eqref{eq:chart}. The transformation between the
canonical and the previously known harmonic coordinates, also left open
in~\cite{Xing:2026rrd}, follows from~\eqref{eq:CS} and~\eqref{eq:chart}. The
Jiang--Lin, or Cook--Scheel, coordinates are obtained from the canonical ones by
the rotation by $\delta$ about the spin axis together with the rescaling of the
cylindrical radius by $1/\cos\delta$ (section~\ref{sec:threecharts}),
\begin{equation}\label{eq:CStocanonical}
(x+iy)_{\rm CS}=\big(1-i\tan\delta(r)\big)\,(x+iy)\,,\qquad
z_{\rm CS}=z\,,\qquad T_{\rm CS}=T\,,
\end{equation}
where $r$ is the function of the canonical point $\bx$ defined
by~\eqref{eq:chart}.

The MPM formalism parametrises the general residual harmonic freedom by the
gauge moments. The results of this section and of
section~\ref{sec:structure} are specific to Kerr: they determine which gauge
moments are compatible with the symmetries of the problem, and give in closed
form the chart in which all of them vanish. The gauge-term-free result
of~\cite{Damgaard:2026kerr} is given by the same chart
(see section~\ref{sec:CS}), in agreement with the comparison of that result with
the canonical metric at fourth PM order in~\cite{Xing:2026rrd}.

\section{Post-Minkowskian expansion}\label{sec:PM}

\subsection{Through third order}

Through order $\GN^3$ the Boyer--Lindquist variables $(R,\theta)$
of~\eqref{eq:chart} coincide with the flat oblate spheroidal coordinates
$(\varrho,\vartheta)$ of the harmonic point $\bx$, eq.~\eqref{eq:oblate}, since
$\cos\delta=1+\mathcal O(\GN^4)$.
Expanding~\eqref{eq:gexact1}--\eqref{eq:gexact3} we find, exactly in the spin,
\begin{align}
\hth^{00}&=\frac{\GN M}{\Sigma}\left[4\varrho+\frac{\GN M\,(7\varrho^2+3a^2)+8\GN^2M^2\varrho}{\varrho^2+a^2}\right]+\mathcal O(\GN^4)\,,\label{eq:h00PM}\\
\hth^{0i}&=\frac{2\GN Ma}{(\varrho^2+a^2)\,\Sigma}\left[\varrho+\GN M+\frac{\GN^2M^2\varrho}{\varrho^2+a^2}\right](\zh\times\bx)^i+\mathcal O(\GN^4)\,,\label{eq:h0iPM}\\
\hth^{ij}&=\frac{\GN^2M^2}{\Sigma}\left[\partial_\varrho x^i\,\partial_\varrho x^j+\frac{a^2\sin^2\vartheta}{\varrho^2+a^2}\,\hat e_\varphi^i\hat e_\varphi^j\right]+\mathcal O(\GN^4)\,,\label{eq:hijPM}
\end{align}
with $\partial_\varrho\bx=\frac{\varrho\sin\vartheta}{\sqrt{\varrho^2+a^2}}\,\evarpi+\cos\vartheta\,\zh$
and $\Sigma=\varrho^2+a^2\cos^2\vartheta$. For $a=0$ these equations reduce to
$\hth^{00}=4\rho+7\rho^2+8\rho^3$ and $\hth^{ij}=\rho^2n^in^j$ with $\rho=\GN
M/|\bx|$, the expansion parameter of~\cite{Mougiakakos:2024nku}. The spatial
components of the inverse gothic metric have no $\GN^3$ term.

The denominators are best read in terms of the complex distance
$\zeta=\varrho-ia\cos\vartheta$ of~\eqref{eq:FTff}. The linear field is a sum of
complex-shifted Coulomb potentials, $\hth^{(1)00}=4\GN M\,\mathrm{Re}(1/\zeta)$,
whose two terms $1/\zeta$ and $1/\bar\zeta$ come from the two exponentials
$e^{\pm\ba\cdot\bk}$ of the form factor $\cosh(\ba\cdot\bk)$. From second order
on, the factor $1/\Sigma=1/(\zeta\bar\zeta)$ mixes them. In momentum space it
comes from the convolution of $e^{\ba\cdot\boldsymbol\ell}$ with
$e^{-\ba\cdot(\bk-\boldsymbol\ell)}$ in the bubble integrals, which, unlike the
products of equal-sign exponentials, do not factor out of the loop integral.

\subsection{Fourth order}

At order $\GN^4$, using~\eqref{eq:deltasmall}, we have
$\delta(R)=G_N^2M^2\delta_2(R)+\mathcal O(\GN^4)$, therefore the factor
$\cos\delta$ contributes and $\arctan(a/\varrho)$ appears, through
\begin{equation}
\delta_2(\varrho)=\frac{1}{2a^2}\left[\arctan\frac{a}{\varrho}-\frac{a\varrho}{\varrho^2+a^2}\right].
\end{equation}
We have for the metric components at this order
\begin{equation}\label{eq:h004}
    \begin{aligned}
\hth^{00}\big|_{\GN^4}&=\GN^4M^4\left[\frac{4(2\varrho^2+a^2)}{(\varrho^2+a^2)^2\,\Sigma}-\frac{a\varrho\sin^2\vartheta}{(\varrho^2+a^2)\,\Sigma}\,\delta_2+\delta_2^2\right],\cr
\hth^{0i}\big|_{\GN^4}&=\frac{2\GN^4M^4a}{(\varrho^2+a^2)^2\,\Sigma}\,(\zh\times\bx)^i\,,
    \end{aligned}
\end{equation}
and, writing 
\begin{equation}
\label{eq:hijdecomp}
\hth^{ij}=\alpha_\perp\delta^{ij}_\perp+\alpha_z\hat z^i\hat z^j+\beta\,x_\perp^ix_\perp^j+\kappa\,(x_\perp^i\hat z^j+\hat z^ix_\perp^j),
\end{equation}
with $\delta_{\perp}^{ij}:=\delta^{ij}-\hat{z}^{i}\hat{z}^{j}$ and $x_{\perp}^{i} := \delta_{\perp}^{ij}x_{j}$.
The coefficients are given by
\begin{equation}\label{eq:hij4}
\begin{aligned}
\alpha_\perp\big|_{\GN^4}&=\GN^4M^4\,\frac{a\sin^2\vartheta\,\big(\varrho(\varrho^2+a^2)\delta_2+a\big)}{(\varrho^2+a^2)^2\,\Sigma}\,,\
&\alpha_z\big|_{\GN^4}&=\GN^4M^4\left[\frac{a\varrho\sin^2\vartheta}{(\varrho^2+a^2)\,\Sigma}\,\delta_2-\delta_2^2\right],\\
\beta\big|_{\GN^4}&=-\GN^4M^4\,\frac{a\big(2\varrho(\varrho^2+a^2)\delta_2+a\big)}{(\varrho^2+a^2)^3\,\Sigma}\,,\
&\kappa\big|_{\GN^4}&=-\GN^4M^4\,\frac{a\cos\vartheta\,\delta_2}{(\varrho^2+a^2)\,\Sigma}\,.
\end{aligned}
\end{equation}
These expressions vanish as $a\to0$, except for the Schwarzschild term
$8\GN^4M^4/|\bx|^4$ in $\hth^{00}$. They agree with the fourth-order canonical
metric of~\cite{Xing:2026rrd}, using the convention $h=-\hth$ for the gothic
perturbation. 

\subsection{Fifth order}

At order $\GN^5$ the inversion of the relation between $\bx$ and $(R,\theta)$
starts to contribute. Since
$\cos\delta=1-\frac12\GN^4M^4\delta_2^2+\mathcal O(\GN^6)$, the
chart~\eqref{eq:chart} gives for the Boyer--Lindquist variables of the harmonic
point $\bx$
\begin{equation}\label{eq:Rtheta4}
\begin{aligned}
R&=\varrho+\GN^4M^4\,\frac{\varrho\,(\varrho^2+a^2)\sin^2\vartheta}{2\Sigma}\,\delta_2^2+\mathcal O(\GN^6)\,,\cr
\theta&=\vartheta+\GN^4M^4\,\frac{(\varrho^2+a^2)\sin\vartheta\cos\vartheta}{2\Sigma}\,\delta_2^2+\mathcal O(\GN^6)\,,
\end{aligned}
\end{equation}
with $\delta_2=\delta_2(\varrho)$. This is the rescaling of the cylindrical
radius by $1/\cos\delta$ at fixed $z$. Acting on the first-order
field~\eqref{eq:h1} it contributes at order $\GN^5$, and the result is again a
polynomial of degree two in $\delta_2$ with rational coefficients,
\begin{equation}\label{eq:h005}
\begin{aligned}
\hth^{00}\big|_{\GN^5}&=\GN^5M^5\bigg[\frac{8\varrho}{(\varrho^2+a^2)^2\,\Sigma}-\frac{4a\varrho^2\sin^2\vartheta}{(\varrho^2+a^2)\,\Sigma^2}\,\delta_2\cr
&\qquad\qquad+\bigg(\frac{4\varrho}{\Sigma}+\frac{2\varrho\,(\varrho^2+a^2)\sin^2\vartheta\,(3a^2\cos^2\vartheta-\varrho^2)}{\Sigma^3}\bigg)\delta_2^2\bigg],\cr
\hth^{0i}\big|_{\GN^5}&=\GN^5M^5\,a\varrho\,\bigg[\frac{2}{(\varrho^2+a^2)^3\,\Sigma}-\frac{2a\varrho\sin^2\vartheta}{(\varrho^2+a^2)^2\,\Sigma^2}\,\delta_2\cr
&\qquad\qquad+\frac{3(\varrho^2+a^2)\cos^2\vartheta-\varrho^2-a^2\cos^4\vartheta}{\Sigma^3}\,\delta_2^2\bigg](\zh\times\bx)^i\,,
\end{aligned}
\end{equation}
while $\hth^{ij}\big|_{\GN^5}=0$, because the spatial components contain only
even powers of $\GN$. The terms quadratic in $\delta_2$ are
$\GN^4M^4\delta_2^2\,\big(1+\frac12\varpi\partial_\varpi\big)$, with the
derivative taken at fixed $z$, acting on $\hth^{(1)00}$ and on the function
multiplying $(\zh\times\bx)^i$ in $\hth^{(1)0i}$. The first term comes from the factor
$\cos^2\delta$ in $\cD$ and the second from the shift~\eqref{eq:Rtheta4}. For
$a\to0$ only the Schwarzschild term $8\GN^5M^5/|\bx|^5$ of $\hth^{00}$
remains. The second term of the expansion~\eqref{eq:delta} of $\delta$, of
order $\GN^4$, first contributes to the metric at order $\GN^6$.

\subsection{Auxiliary field}

The auxiliary currents follow from~\eqref{eq:Aonshell} and~\eqref{eq:Ydef},
$Y=A-A_{\Lin}[\hth]$, evaluated on the chart~\eqref{eq:chart}. 
At order $\GN^2$ and $a=0$ we find the Schwarzschild result~\cite{Mougiakakos:2024nku}
\begin{equation}\label{eq:Y2}
Y^{(2)\lambda}{}_{\mu\nu}=-2\,\partial^\lambda\rho^2\,\big(3\,\delta^0_\mu\delta^0_\nu+\eta_{\mu\nu}\big)-8\,\delta^\lambda_0\,\delta^0_{(\mu}\partial_{\nu)}\rho^2\,,\qquad \rho=\frac{\GN M}{|\bx|}\,,
\end{equation}
i.e.\ $Y^{(2)i}{}_{00}=Y^{(2)0}{}_{0i}=8\GN^2M^2x^i/|\bx|^4$ and
$Y^{(2)i}{}_{jk}=4\GN^2M^2x^i\delta_{jk}/|\bx|^4$. The spin-dependent terms
start at order $\GN^2a$ and are generated in the same way.

\section{Explicit checks}\label{sec:checks}

We implemented the recursion~\eqref{eq:recY}--\eqref{eq:recJ} in position space.
Fields are represented as sums $\sum_{n,p}\GN^nM^na^p\,P_{n,p}(\bx)/|\bx|^{m}$
with homogeneous polynomials $P_{n,p}$. Sources are decomposed into harmonic
polynomials and eq.~\eqref{eq:invlap} is applied, with an explicit resonance
check. The cubic recursion reproduces the expansion of the exact
metric~\eqref{eq:gexact1}--\eqref{eq:gexact3} for all ten components through
$\GN^5$ and spin order $a^{12}$ in exact rational arithmetic. The lowest-order
coefficients are listed in appendix~\ref{app:coefficients}. At each computed
order the residuals of eqs.~\eqref{eq:Aeq} and~\eqref{eq:geq} vanish, the
harmonic condition holds, and no resonance is encountered. The closed forms of
section~\ref{sec:PM}, including~\eqref{eq:h005}, were compared with the
recursion through order $a^{12}$ and with the Taylor coefficients of the exact
metric at generic points.

The source-side checks verify that the linear worldline coupling
of~\eqref{eq:Swl} reproduces the Kerr stress tensor~\eqref{eq:Trest} through
spin order $a^{13}$. We also checked the bulk-action decomposition, inverse
kinetic operator, Euler--Lagrange equations and source term in $D=3,\ldots,6$.
These algebraic checks are independent of the metric comparison.

As an independent formulation, the Landau--Lifshitz relaxed equations of
appendix~\ref{app:LL} reproduce the same exterior solution through the same
orders. Continued through order $\GN^8$, the same computation is the iteration
of section~\ref{sec:canonical}, no resonance is met, and its solution is
divergence-free and coincides with that of the cubic recursion. The exact metric
expressions agree with a direct tensor-density transformation of the
Boyer--Lindquist metric to 36 significant digits at generic points. Finally, we
compared the closed forms of section~\ref{sec:PM} with the canonical MPM metric
of~\cite{Xing:2026rrd}, which they reproduce at each order through fourth PM
order, for all components and exactly in the spin (section~\ref{sec:canonical}),
and with the gauge-term-free recursion of~\cite{Damgaard:2026kerr}. The negative
controls are also satisfied: the recursion does not generate the time-odd
Cook--Scheel term, and replacing $\cosh$ by $\cos$ in the seed spoils the
agreement already at order $\GN^2a^2$. The ranges and outcomes are summarised in
table~\ref{tab:checks}. A Python implementation of the recursion and of the
checks listed above, together with the spin-expanded coefficients, is available
in the accompanying Zenodo archive~\cite{Sadu:2026code}. The repository includes
the independent comparisons with the Landau--Lifshitz recursion, with the
canonical MPM metric and with the gauge-term-free recursion
of~\cite{Damgaard:2026kerr} discussed in section~\ref{sec:CS}.

\begin{table}[htb]
\centering
\small
\begin{tabularx}{\textwidth}{@{}Xll@{}}
\toprule
check & range & result\\
\midrule
cubic recursion \eqref{eq:recY}--\eqref{eq:recJ} vs.\ \eqref{eq:chart} &
$n\le5$, $p\le12$, all components & exact agreement\\
residuals of \eqref{eq:Aeq}, \eqref{eq:geq}; $\partial_\mu\hth^{\mu\nu}$ & $n\le5$, $p\le12$ & zero\\
resonances (logarithms) & $n\le5$, $p\le12$ & none\\
identities of the cubic formulation (appendix~\ref{app:cubic}) & $D=3,\dots,6$ & identities hold\\
worldline action \eqref{eq:Swl} vs.\ \eqref{eq:Trest} & $p\le13$ & exact agreement\\
Landau--Lifshitz recursion vs.\ \eqref{eq:chart} & $n\le5$, $p\le12$ & exact agreement\\
\eqref{eq:gexact1}--\eqref{eq:gexact3} vs.\ direct transformation & generic points & $<10^{-36}$\\
\eqref{eq:Aonshell} vs.\ recursion & $n\le5$, $p\le12$ & exact agreement\\
\eqref{eq:h1}, \eqref{eq:h00PM}--\eqref{eq:hij4} vs.\ \cite{Xing:2026rrd} & $n\le4$, all components & exact agreement\\
\eqref{eq:h004}--\eqref{eq:hij4} vs.\ \cite{Damgaard:2026kerr} without gauge terms & $n=4$, $p\le20$ & exact agreement\\
MPM iteration \eqref{eq:MPMalgorithm} vs.\ cubic recursion
& $n\le8$, $p\le12$
& $v_{(n)}=0$, exact agreement\\
\bottomrule
\end{tabularx}
\caption{Summary of the explicit checks; $n$ is the order in $\GN$ and $p$ the order in the spin.}
\label{tab:checks}
\end{table}

\paragraph{Dimensional regularisation.} If the recursion is performed in
momentum space with dimensional regularisation, as
in~\cite{Mougiakakos:2020laz,Damgaard:2024fqj,Mougiakakos:2024nku,Damgaard:2026kerr},
the residual homogeneous ambiguity relevant to the four-dimensional limit is the
one singled out by Theorem~\ref{thm:homog}, namely the spatial components at
order $\GN^3$ with $\ell=p+2$. In the corresponding resonant channel, the
denominator in the $D$-dimensional analogue of~\eqref{eq:invlap} has a simple
zero at $D=4$. For Schwarzschild the corresponding source vanishes quadratically
in $D-4$ (see appendix~\ref{app:tangherlini}), so the $D\to4$ limit is the
standard chart, in agreement with~\cite{Mougiakakos:2024nku}. For Kerr the
answer depends on how the spinning source is continued to $D$ dimensions. A
finite remainder would only amount to a residual gauge transformation at order
$\GN^3$. With the prescription of~\cite{Damgaard:2026kerr} the result agrees
with the chart~\eqref{eq:chart} through fourth order as explained in
section~\ref{sec:CS}. Within the source continuation and boundary prescription
used in~\cite{Damgaard:2026kerr}, such a remainder is fixed to zero in the
third-order spatial sector.  As shown in this paper, the all-order
identification then follows from Theorem~\ref{thm:unique}. The other
difficulties discussed in~\cite{Damgaard:2026kerr}, Fourier transforms that are
singular in three dimensions and the failure of $r^2=x^2+y^2+z^2$ in
$d=3-2\epsilon$, do not arise for a stationary source treated in position space.
There~\eqref{eq:invlap} is algebraic, and its only possible singularity is the
resonance of Theorem~\ref{thm:homog}, which does not occur.

\section{Discussion}\label{sec:discussion}

We have classified the residual harmonic-coordinate freedom in the perturbative
Kerr solution and identified the representative selected by the Kerr multipole
source and the stated boundary prescription. The main result is the restriction
of possible decaying homogeneous solutions to the exceptional spatial sector at
third PM order, together with the exact chart~\eqref{eq:chart} and its inverse
gothic metric~\eqref{eq:gexact1}--\eqref{eq:gexact3}. The chart is time-reversal
covariant and is constructed from the real combination of the two
complex-conjugate radial harmonic functions. Its expansion agrees with the
harmonic-gauge recursion without the additional order-$\GN^2$ gauge term,
including the fourth-PM coefficients available in~\cite{Damgaard:2026kerr}. The
Cook--Scheel chart is obtained by a definite residual harmonic transformation.
The chart is the canonical harmonic chart of the MPM formalism to all PM orders,
and its expansion reproduces the canonical metric of~\cite{Xing:2026rrd} through
fourth PM order for all components. The MPM formalism already parametrises
general harmonic-gauge freedom by gauge-moment families. Our contribution is the
Kerr-sector restriction and selection of admissible homogeneous modes, their
identification with gauge moments, and the closed form of the canonical chart.
It is not a first general classification of residual harmonic-coordinate
freedom.

The all-order argument has two logically distinct ingredients. First, rotational
covariance, parity and time reversal constrain the allowed tensor structures and
restrict decaying homogeneous solutions to the spatial components at third PM
order. Second, the explicit harmonic Kerr chart has no spatial contribution in
that exceptional sector, and the recursion is defined with the same boundary
condition. The absence of logarithmic terms is a separate consequence of the
existence of this exact solution whose PM expansion contains no logarithmic
terms. Symmetry alone does not uniquely determine the chart. The uniqueness
statement identifies the solution generated by the recursion with the known
exact Kerr geometry under the stated source and boundary prescriptions.

The cubic variables $(\hth,Y)$ provide an efficient computational representation
of the Einstein recursion. The Landau--Lifshitz relaxed equations give an
independent formulation and reproduce the same result. The exact result is
checked by an explicit recursion through $\GN^5$ and spin order $a^{12}$, by an
independent Landau--Lifshitz computation, and by comparison with the available
fourth-order canonical MPM expansion and with the fourth-order result
of~\cite{Damgaard:2026kerr}. These checks validate the closed-form
identification while keeping the finite-order calculations distinct from the
all-order uniqueness argument. The extended Polyakov
construction~\cite{BjerrumBohr:2026polyakov} is complementary at the level of
the Kerr three-point source, while the third-PM scattering analysis
of~\cite{BjerrumBohr:2026resum} addresses spin-resummed observables rather than
the stationary harmonic metric. Neither work removes the need to specify the
exceptional homogeneous sector in the present coordinate-reconstruction problem.

Several extensions are worth pursuing. A first direction is to allow general
spin-induced multipole coefficients in the worldline source, rather than fixing
them to their Kerr values. This would provide a systematic map from worldline
Wilson coefficients to the asymptotic mass and current multipoles of a generic
stationary spinning object. At linear order the map is eq.~\eqref{eq:Swllin},
and the non-linear metric of stationary sources with general multipoles was
studied at second order in~\cite{Damgaard:2026star}. A second direction is the
coupled Einstein--Maxwell problem for Kerr--Newman, where the photon and
graviton source amplitudes and their nonlinear interactions must be treated
together. The existence of known harmonic charts provides a useful target but
does not by itself establish the corresponding recursion. Higher-dimensional
Myers--Perry geometries~\cite{Myers:1986un} offer a further test, particularly
because the classification of homogeneous solutions and the
dimensional-regularisation resonances depend on the spacetime dimension.
Finally, the all-order harmonic Kerr metric provides a fixed gauge background
for amplitude-based probe-scattering~\cite{Bautista:2021wfy} and self-force
calculations, allowing direct comparisons between metric and amplitude methods.

\acknowledgments
A.S.\ thanks the Institut de Physique Th\'eorique for its hospitality during her
M1 internship. The work of P.V.{} was supported by the Agence Nationale de la
Recherche (ANR) under the grant Observables (ANR-24-CE31-7996). 
We acknowledge the use of Claude for language editing and assistance in
developing the Python codes. The authors checked and validated the implementation
and all results.

\appendix

\section{Landau--Lifshitz form of the relaxed equations}\label{app:LL}

As an independent check we also solved the relaxed Einstein equations in the
Landau--Lifshitz form~\cite{Landau:1975pou,Poisson:2014}, which in harmonic
gauge and outside the source read
\begin{equation}
\Box\,\hth^{\alpha\beta}=\hth^{\mu\nu}\partial_\mu\partial_\nu\hth^{\alpha\beta}-\partial_\mu\hth^{\alpha\nu}\partial_\nu\hth^{\beta\mu}-16\pi\GN(-g)\,t_{LL}^{\alpha\beta}\,,\label{eq:LLwave}
\end{equation}
with $\hth^{\alpha\beta}=\eta^{\alpha\beta}-\gth^{\alpha\beta}$ and
\begin{align}
16\pi\GN(-g)\,t_{LL}^{\alpha\beta}&=\frac12g^{\alpha\beta}g_{\lambda\mu}\partial_\rho\gth^{\lambda\nu}\partial_\nu\gth^{\mu\rho}-g^{\alpha\lambda}g_{\mu\nu}\partial_\rho\gth^{\beta\nu}\partial_\lambda\gth^{\mu\rho}-g^{\beta\lambda}g_{\mu\nu}\partial_\rho\gth^{\alpha\nu}\partial_\lambda\gth^{\mu\rho}\nonumber\\
&\quad+g_{\lambda\mu}g^{\nu\rho}\partial_\nu\gth^{\alpha\lambda}\partial_\rho\gth^{\beta\mu}\nonumber\\
&\quad+\frac18\left(2g^{\alpha\lambda}g^{\beta\mu}-g^{\alpha\beta}g^{\lambda\mu}\right)\left(2g_{\nu\rho}g_{\sigma\tau}-g_{\rho\sigma}g_{\nu\tau}\right)\partial_\lambda\gth^{\nu\tau}\partial_\mu\gth^{\rho\sigma}\,,
\end{align}
where the terms proportional to $\partial_\mu\gth^{\mu\nu}$ have been dropped.
Here $g^{\mu\nu}=\gth^{\mu\nu}/\sqrt{-g}$ and
$g_{\mu\nu}=\sqrt{-g}\,(\gth^{-1})_{\mu\nu}$ with $-g=-\det\gth^{\mu\nu}$ in
four dimensions. The two forms of the field equations differ by terms that
vanish in harmonic gauge. With the same source data and inverse-Laplacian
boundary prescription, they give the same exterior solution.

\section{Harmonic radius in $D$ dimensions}\label{app:tangherlini}

For the Schwarzschild--Tangherlini metric~\cite{Tangherlini:1963bw} with
$f(r)=1-\mu/r^{D-3}$ (so that $\mu=2\GN M$ in four dimensions), the harmonic radial function $\varrho(r)$ satisfies
$\partial_r\big(r^{D-2}f(r)\partial_r\varrho(r)\big)=(D-2)\,r^{D-4}\varrho(r)$.
The solution growing at infinity has the post-Minkowskian expansion
$\varrho(r)=r\sum_{k\ge0}c_k\,(\mu r^{3-D})^k$ with $c_0=1$ and, writing
$\epsilon=D-3$,

\begin{equation}
c_k=\frac{\big(1-(k-1)\epsilon\big)^2}{k\,\epsilon\,\big((k-1)\epsilon-2\big)}\;c_{k-1}\,.
\end{equation}
At $D=4$ the factor $1/((k-1)\epsilon-2)$ has a pole at $k=3$, the resonance of
Theorem~\ref{thm:homog}. However $c_2\propto(D-4)^2$, so that $c_3\propto D-4$
and $c_{k\ge2}\to0$ as $D\to4$. The limit is $\varrho=r-\GN M$, the standard
harmonic chart with no admixture of the decaying solution $(r-\GN M)\log(1-2\GN
M/r)+2\GN M$.

\begin{table}[t]
\centering
\renewcommand{\arraystretch}{1.2}
\begin{tabular}{@{}cccc@{}}
\toprule
$n$ & $h_{n,0}$ & $h_{n,2}$ & $h_{n,4}$\\
\midrule
1 & $4$ & $2(1-3\xi^2)$
  & $\tfrac{1}{2}(35\xi^4-30\xi^2+3)$\\
2 & $7$ & $3-14\xi^2$
  & $3(14\xi^4-10\xi^2+1)$\\
3 & $8$ & $4(1-5\xi^2)$
  & $3(21\xi^4-14\xi^2+1)$\\
4 & $8$ & $\tfrac{1}{9}(34-213\xi^2)$
  & $\tfrac{1}{15}(1175\xi^4-696\xi^2+57)$\\
5 & $8$ & $\tfrac{2}{9}(13-119\xi^2)$
  & $\tfrac{2}{45}(1980\xi^4-963\xi^2+29)$\\
\bottomrule
\end{tabular}
\caption{Temporal coefficients $h_{n,p}$ in the spin expansion
\eqref{eq:spinexp} through order $\GN^5$, for $p=0,2,4$.
Here $\xi=\zh\cdot\bn$.}
\label{tab:coefficients-h}
\end{table}

\section{Spin expansion of the post-Minkowskian coefficients}\label{app:coefficients}

The recursion of section~\ref{sec:checks} produces the metric as a double
expansion in $\GN$ and in the spin. With $\bn=\bx/|\bx|$, $\xi:=\zh\cdot\bn$
and $\rho=\GN M/|\bx|$ we write
\begin{equation}\label{eq:spinexp}
\hth^{00}=\sum_{n\ge1}\sum_{p\ge0}\rho^n\Big(\frac{a}{|\bx|}\Big)^{p}\,h_{n,p}(\xi)\,,\qquad
\hth^{0i}=(\zh\times\bn)^i\sum_{n\ge1}\sum_{p\ge1}\rho^n\Big(\frac{a}{|\bx|}\Big)^{p}\,w_{n,p}(\xi)\,,
\end{equation}
and, for the functions of the decomposition~\eqref{eq:hijdecomp} of the spatial
components,  
\begin{equation}\label{eq:spinexpij}
\big(\alpha_\perp,\,\alpha_z,\,|\bx|^2\beta,\,|\bx|\,\kappa\big)=\sum_{n\ge2}\sum_{p\ge0}\rho^n\Big(\frac{a}{|\bx|}\Big)^{p}\,\big(\alpha^\perp_{n,p},\,\alpha^z_{n,p},\,\beta_{n,p},\,\kappa_{n,p}\big)(\xi)\,.
\end{equation}
The coefficients are polynomials in $\xi$ with rational coefficients, although
the closed forms~\eqref{eq:h004}--\eqref{eq:hij4} and~\eqref{eq:h005} contain
$\arctan(a/\varrho)$. By time reversal, condition~(iii) of
section~\ref{sec:structure}, $h_{n,p}$ and the spatial coefficients vanish for
odd $p$ and $w_{n,p}$ vanishes for even $p$. The coefficients of the
spatial-spatial components $\alpha^\perp_{n,p}$, $\alpha^z_{n,p}$, $\beta_{n,p}$
and $\kappa_{n,p}$ also vanish for odd $n$, since the spatial components contain
only even powers of $\GN$.
Table~\ref{tab:coefficients-h}--\ref{tab:coefficients-tensor} give the three
lowest spin orders through order $\GN^5$. For $p=0$ they are the Schwarzschild
values $\hth^{00}=4\rho+7\rho^2+8\rho^3+8\rho^4+8\rho^5$ and
$\hth^{ij}=\rho^2n^in^j$. The entries $w_{n,1}=2$ are the expansion of the term
linear in the spin, which is $\hth^{0i}=2\GN
Ma\,(\zh\times\bx)^i/\big(|\bx|^2(|\bx|-\GN M)\big)+\mathcal O(a^3)$ to all
orders in $\GN$. The coefficients through order $a^{12}$ are given in the file
\texttt{kerr-\allowbreak harmonic-\allowbreak coefficients.m} of the
repository~\cite{Sadu:2026code}.

\begin{table}[t]
\centering
\renewcommand{\arraystretch}{1.2}
\begin{tabular}{@{}cccc@{}}
\toprule
$n$ & $w_{n,1}$ & $w_{n,3}$ & $w_{n,5}$\\
\midrule
1 & $2$ & $1-5\xi^2$
  & $\tfrac{3}{4}(21\xi^4-14\xi^2+1)$\\
2 & $2$ & $2(1-3\xi^2)$
  & $2(10\xi^4-8\xi^2+1)$\\
3 & $2$ & $1-7\xi^2$
  & $\tfrac{3}{4}(33\xi^4-18\xi^2+1)$\\
4 & $2$ & $2(1-4\xi^2)$
  & $2(15\xi^4-10\xi^2+1)$\\
5 & $2$ & $\tfrac{2}{9}(1-36\xi^2)$
  & $\tfrac{1}{180}(5005\xi^4-1518\xi^2-63)$\\
\bottomrule
\end{tabular}
\caption{Mixed time-space coefficients $w_{n,p}$ in the spin
expansion~\eqref{eq:spinexp} through order $\GN^5$, for
$p=1,3,5$. Here $\xi=\zh\cdot\bn$.}
\label{tab:coefficients-w}
\end{table}

\begin{table}[H]
\centering
\renewcommand{\arraystretch}{1.2}
\begin{tabular}{@{}cccc@{}}
\toprule
 & $p=0$ & $p=2$ & $p=4$\\
\midrule
$\alpha^\perp_{2,p}$ & $0$ & $1-\xi^2$
  & $(1-\xi^2)(1-4\xi^2)$\\
$\alpha^z_{2,p}$ & $\xi^2$ & $\xi^2(2-3\xi^2)$
  & $\xi^2(10\xi^4-12\xi^2+3)$\\
$\beta_{2,p}$ & $1$ & $-(1+3\xi^2)$
  & $10\xi^4-1$\\
$\kappa_{2,p}$ & $\xi$ & $\xi(1-3\xi^2)$
  & $\xi(10\xi^4-8\xi^2+1)$\\
\addlinespace
$\alpha^\perp_{4,p}$ & $0$ & $\tfrac{4}{3}(1-\xi^2)$
  & $\tfrac{1}{15}(1-\xi^2)(19-100\xi^2)$\\
$\alpha^z_{4,p}$ & $0$ & $\tfrac{1}{9}(2-3\xi^2)$
  & $\tfrac{1}{15}(25\xi^4-24\xi^2+3)$\\
$\beta_{4,p}$ & $0$ & $-\tfrac{5}{3}$
  & $\tfrac{1}{15}(125\xi^2-23)$\\
$\kappa_{4,p}$ & $0$ & $-\tfrac{1}{3}\xi$
  & $\tfrac{1}{15}\xi(25\xi^2-9)$\\
\bottomrule
\end{tabular}
\caption{Tensorial coefficients in the spatial spin expansion
\eqref{eq:spinexpij}, for $n=2,4$ and $p=0,2,4$.
Here $\xi=\zh\cdot\bn$.}
\label{tab:coefficients-tensor}
\end{table}

\section{The cubic formulation of the action}\label{app:cubic}

This appendix collects the properties of the cubic formulation of the
Einstein--Hilbert action~\cite{Cheung:2017kzx,Mougiakakos:2024nku} that are used in
section~\ref{sec:cubic}: the identity behind the field equation~\eqref{eq:Aeq},
the form of the action in the variables $(\hth,Y)$ and its Feynman rules, and the
derivation of the field equations~\eqref{eq:Yeq} and~\eqref{eq:heq}. These
identities are checked in exact arithmetic in $D=3,\dots,6$ in the accompanying
code~\cite{Sadu:2026code}.

\paragraph{Effective action.} The effective action is the sum of the bulk action
and of the worldline action of the body,
\begin{equation}\label{eq:Seff}
S_{\rm eff}=\int d^Dx\;\mathcal L_{\rm EH}\;+\;S_{\rm wl}[z,S;g]\,,
\end{equation}
where $\mathcal L_{\rm EH}$ is given in~\eqref{eq:action}, and $S_{\rm wl}$,
given in~\eqref{eq:Swl}, depends on the trajectory $z^\mu(\tau)$ and on the spin
tensor $S^{\mu\nu}(\tau)$ of the body, and on the metric $g_{\mu\nu}$, but not
on the auxiliary field $A$.

\paragraph{Field equation of the auxiliary field.} With the notation
\begin{equation}
    A\cdot X:= A^\lambda{}_{\mu\nu}\,X_\lambda{}^{\mu\nu}\,,
\end{equation}
the bilinear form~\eqref{eq:Qdef} and the map~\eqref{eq:Aeq} are related by
\begin{equation}\label{eq:QE}
A\cdot\cE(B;\gth)=2\,Q(A,B;\gth)\,,
\end{equation}
valid for any symmetric $\gth^{\mu\nu}$. The variation of~\eqref{eq:action}
with respect to $A$ therefore gives~\eqref{eq:Aeq}. In particular the constant
map $\cE_0=\cE(\,\cdot\,;\eta)$ is symmetric, $A\cdot\cE_0(B)=B\cdot\cE_0(A)$,
and so is its inverse~\eqref{eq:E0inv}.

\paragraph{Bulk action.} Since
\begin{equation}
A\cdot\partial\gth=-A\cdot\partial\hth=A\cdot\cE_0(A_{\Lin})=2Q(A,A_{\Lin};\eta)\,,
\end{equation}
the first two terms of the Einstein-Hilbert action~\eqref{eq:action} become,
with the shift~\eqref{eq:Ydef},
\begin{equation}
-Q(Y,Y;\eta)+Q(A_{\Lin},A_{\Lin};\eta)+Q(A,A;\hth).
\end{equation}
Moreover $Q(A_{\Lin},A_{\Lin};\eta)=-\frac12A_{\Lin}\cdot\partial\hth$, which
combines with the gauge-fixing term into the de~Donder kinetic term. Up to a
total derivative, the bulk Lagrangian is
\begin{equation}\label{eq:Lbulk}
16\pi\GN\,\mathcal L_{\rm EH}=\frac14\,\hth^{\mu\nu}\,\Box\,(P\hth)_{\mu\nu}-Q(Y,Y;\eta)+Q(A,A;\hth)\,,
\end{equation}
with the trace reversal $P$ of~\eqref{eq:Pdef}.

\paragraph{Feynman rules.} The Feynman rules derived from~\eqref{eq:Lbulk} are
given in table~\ref{tab:rules}:
\begin{enumerate}
\item[(a)] \emph{The quadratic part is diagonal.} The shift~\eqref{eq:Ydef}
completes the square in $A$, so that the metric perturbation $\hth$ and the
auxiliary field perturbation $Y$ do not mix at quadratic order. The graviton has
the de~Donder kinetic term and the quadratic term of the auxiliary field $Y$
has no derivatives.
\item[(b)] \emph{Three cubic vertices.} The interactions are the expansion of
the last term of~\eqref{eq:Lbulk},
\begin{equation}\label{eq:vertices}
Q(A,A;\hth)=\underbrace{Q(A_{\Lin},A_{\Lin};\hth)}_{\hth\hth\hth}+\underbrace{2\,Q(A_{\Lin},Y;\hth)}_{\hth\hth Y}+\underbrace{Q(Y,Y;\hth)}_{\hth YY}\,,
\end{equation}
with two, one and zero derivatives respectively (recall that $A_{\Lin}\sim
\partial\hth$ in~\eqref{eq:Alin}). There are no other interactions since there
is no $YYY$ vertex, no vertex from the gauge-fixing term, and no quartic or
higher vertices. The undifferentiated leg of each vertex is the graviton
$\hth^{\rho\sigma}$ contracted with $(AA)_{\rho\sigma}$.
\item[(c)] \emph{Propagators.} The graviton propagator is given
by~\eqref{eq:props}. For a source term $\int d^Dx\,Y\cdot K$, the
auxiliary field is $Y=16\pi\GN\,\cE_0^{-1}K$, without dependence on the momentum
because $Y$ does not propagate, and its exchange is a contact interaction only.
\item[(d)] \emph{Relation with the Einstein--Hilbert action.} On shell, the
auxiliary field $Y$ is the part of the connection~\eqref{eq:Aonshell} that is
non-linear in $\hth$. Its field equation~\eqref{eq:Yeq} is algebraic, and can
be written as $\cE(Y;\gth)=\cE(A_{\Lin}[\hth];\hth)$. Its solution is
non-polynomial in $\hth$, and substituting it into~\eqref{eq:Lbulk} gives back
the Einstein--Hilbert action in gothic variables, with its infinite tower of
vertices. Diagrammatically, chains of contact $Y$-exchanges between the cubic
vertices collapse into the higher-point vertices of the Einstein--Hilbert
action. The cubic formulation trades them for three cubic vertices and a
non-propagating field. We refer to~\cite{Mougiakakos:2024nku} for a detailed
discussion applied to the Schwarzschild case.
\end{enumerate}

\begin{table}[t]
\centering
\renewcommand{\arraystretch}{1.7}
\begin{tabular}{@{}lllc@{}}
\toprule
& term in $16\pi\GN\,\mathcal L$ & propagator, vertex & derivatives\\
\midrule
graviton $\hth^{\mu\nu}$ & $\frac14\,\hth^{\mu\nu}\,\Box(P\hth)_{\mu\nu}$ & \lineG\quad$32\pi\GN\,P^{-1}/\bk^2$ & 2\\
auxiliary field $Y^\lambda{}_{\mu\nu}$ & $-Q(Y,Y;\eta)$ & \lineY\quad$16\pi\GN\,\cE_0^{-1}$ \ (contact) & 0\\
$\hth\hth\hth$ vertex & $Q(A_{\Lin},A_{\Lin};\hth)$ & \vertexS{grav}{grav}{grav} & 2\\
$\hth\hth Y$ vertex & $2\,Q(A_{\Lin},Y;\hth)$ & \vertexS{grav}{grav}{aux} & 1\\
$\hth YY$ vertex & $Q(Y,Y;\hth)$ & \vertexS{grav}{aux}{aux} & 0\\
worldline source & $8\pi\GN\,\hth^{\mu\nu}(PT)_{\mu\nu}$ & $j=\frac12\,PT$ & --\\
\bottomrule
\end{tabular}
\caption{Feynman rules for the effective action~\eqref{eq:Lbulk}, \eqref{eq:Swlexp}
in the variables $(\hth,Y)$, with $A_{\Lin}=A_{\Lin}[\hth]$ given
in~\eqref{eq:Alin}. The propagators are the static Green's functions
of~\eqref{eq:props}, and each vertex carries a factor $1/(16\pi\GN)$. The
exterior recursion is defined by the linear Kerr multipoles and the stated
boundary prescription (section~\ref{sec:nonlinwl}). We follow the notation of~\cite{Mougiakakos:2024nku} for Feynman graphs.}
\label{tab:rules}
\end{table}

\paragraph{Field equations.} The Euler--Lagrange equations of~\eqref{eq:Lbulk}
and~\eqref{eq:Swlexp} follow from the identity~\eqref{eq:QE} and from the
symmetry of $\cE_0^{-1}$. The variation with respect to $Y$
gives~\eqref{eq:Yeq}. In the variation with respect to $\hth$ at fixed $Y$,
$\hth$ appears explicitly in $Q(A,A;\hth)$ and through $A_{\Lin}[\hth]$. Using
\begin{equation}
A_{\Lin}[\delta\hth]\cdot\cE(A;\hth)=-\partial(\delta\hth)\cdot\cE_0^{-1}[\cE(A;\hth)]
\end{equation}
and integrating by parts, we get
\begin{equation}
\frac12\,\Box(P\hth)_{\mu\nu}+(AA)_{\mu\nu}+\partial_\lambda\big(\cE_0^{-1}[\cE(A;\hth)]\big)^\lambda{}_{\mu\nu}=-8\pi\GN\,(PT)_{\mu\nu}\,.
\end{equation}
By~\eqref{eq:Yeq}, the last term on the left-hand side is $\partial_\lambda
Y^\lambda{}_{\mu\nu}$, which gives~\eqref{eq:heq}. Since
\begin{equation}\label{eq:kinetic}
\partial_\lambda A_{\Lin}[\hth]^\lambda{}_{\mu\nu}-\eta_{\sigma(\mu}\partial_{\nu)}\partial_\rho\gth^{\rho\sigma}=\frac12\,\Box(P\hth)_{\mu\nu}\,,
\end{equation}
eq.~\eqref{eq:heq} is the field equation~\eqref{eq:geq} of $\gth$ in the
presence of the source.

\bibliographystyle{JHEP}

\providecommand{\href}[2]{#2}\begingroup\raggedright\endgroup
\end{document}